\RequirePackage{fix-cm}
\documentclass[smallextended]{svjour3}
\usepackage[normalem]{ulem}
\usepackage{epsfig}
\newcommand{\braket}[2]{\langle#1|#2\rangle}

\newcommand{\ket}[1]{|#1\rangle}
\newcommand{\be}{\begin{equation}}
\newcommand{\ee}{\end{equation}}
\newcommand{\ben}{\begin{enumerate}}
\newcommand{\een}{\end{enumerate}}
\newcommand{\bq}{\begin{quote}}
\newcommand{\eq}{\end{quote}}
\newcommand{\hA}{{\bf\hat A}}
\newcommand{\hB}{{\bf\hat B}}
\newcommand{\hE}{{\bf\hat E}}
\newcommand{\hF}{{\bf\hat F}}
\newcommand{\hP}{{\bf\hat P}}

\newcommand{\hW}{{\bf\hat W}}
\newcommand{\hO}{{\bf\hat O}}

\newcommand{\hPi}{{\bf\hat\Pi}}
\newcommand{\hI}{{\bf\hat I}}
\newcommand{\cH}{{\cal H}}

\newcommand{\cP}{{\cal P}}
\newcommand{\cS}{{\cal S}}
\newcommand{\cU}{{\cal U}}
\newcommand{\cL}{{\cal L}}
\newcommand{\cV}{{\cal V}}
\newcommand{\ketbra}[2]{\vert#1\rangle\langle#2\vert}
\newcommand{\sandwich}[3]{\langle#1\vert#2\vert#3\rangle}
\newcommand{\Tr}{\hbox{Tr}}
\begin{document}

\title{QBism: A Critical Appraisal}%
\author{U.J. Mohrhoff}

\institute{Ulrich Mohrhoff \at
              Sri Aurobindo International Centre of Education\\
              605002 Pondicherry India\\
              \email{ujm@auromail.net}}
\date{}
\maketitle
\vspace{-1in}
\noindent \textbf{Note: Neither the original version of this paper nor this replacement reflects the author's most recent views. Readers may want to consult the Appendix for the history of this work and the reasons for this replacement.}

\vspace{0.5in}
\begin{abstract}
QBism (short for Quantum Bayesianism) is a novel interpretation of quantum mechanics. With its radical emphasis on the subject, QBism provides a welcome corrective to popular misrepresentations of the epistemological reflections of Niels Bohr, while Bohr, rightly understood, can take the edge off QBism's radical subjectivism. Together, they form a winning combination, the key to which is Schr\"odinger's concept of objectivation, which has its roots in Kant's theory of science. Objectivation contrasts with objectification, the demonstrably futile attempt to make measurements yield outcomes in interpretations that reify a calculational device. Apart from providing an in-depth discussion of QBism, Kant, and Bohr, the paper addresses the issue of nonlocality, the scenario of Wigner's friend, Bell's shifty split, and the problem of the now. It also outlines an ontological interpretation that incorporates the insights of QBism and Bohr.
\end{abstract}


\section{Introduction}
The beginning of the 21st Century saw the launch of a novel epistemic interpretation of quantum mechanics, by {Carlton Caves}, {Chris Fuchs}, and {Ruediger Schack}.\cite{CFS2002} Initially conceived as an extended personalist Bayesian theory of probability called {``Quantum Bayesianism},'' it has since been re-branded as ``QBism,'' the term Mermin\cite{MerminQBnotCop} prefers, considering it ``as big a break with 20th century ways of thinking about science as Cubism was with 19th century ways of thinking about art.'' The big break lies not in the emphasis that the mathematical apparatus of quantum mechanics is a probability calculus---a point of view I have been defending since about the time QBism came along---but in this \emph{plus} a radically subjective Bayesian interpretation of probability \emph{plus} a radically subjective interpretation of the events to which, and on the basis of which, probabilities are assigned. As an interpretation of the fundamental theoretical framework of contemporary physics, QBism belongs in a category of its own. Any attempt to compare it with a \hbox{``$\Psi$-ontological''} interpretation is bound to be prejudiced by the very expectations that may be responsible for the theory's imperviousness to making sense. In this way and others it has more in common with the epistemological considerations of Niels Bohr than QBists seem prepared to envisage.

According to Mermin,\cite{MerminQBnotCop} the four stages of acceptance of a radical new idea are: (1) It's nonsense%
\footnote{I too started out at stage (1), albeit for reasons vastly different from those proffered by quantum-state realists.}; 
(2) It's well known; (3) It's trivial; (4) I thought of it first. By now Stage (2) appears to be well underway. A growing minority of philosophically-minded physicists are beginning to maintain that there is nothing very new in QBism, that it is in fact just another version of the Copenhagen interpretation.%
\footnote{The term ``Copenhagen interpretation'' appeared in print for the first time in the 1950s, in an essay by Werner Heisenberg.\cite{Heisenberg1955} It was the decade in which David Bohm\cite{BohmHV} presented his hidden-variables interpretation and Hugh Everett III\cite{EverettRSF} put forward his relative-state interpretation. Heisenberg, entering the fray, then argued that the Copenhagen interpretation was the only viable interpretation, thereby transforming Bohr's views into another ontological interpretation of a mathematically formulated theory. In this form Bohr's epistemological reflections could not but seem amateurish, ad hoc, or worse.}
Here I have nothing further to say about ``the'' Copenhagen interpretation, which can mean entirely different things to different people. My interest in much of the present paper lies in the relationship between QBism and Bohr. QBism and Bohr complement each other: QBism, with its radical emphasis on the subject, provides a welcome corrective to popular misrepresentations of Bohr's views, and Bohr, rightly understood, can take the edge off QBism's radical subjectivism.

My other interest in the present paper, apart from offering a critical assessment of QBism, lies in presenting (in outline) an ontology that incorporates the insights of QBism and Bohr. Although my assessment of QBism has changed from harshly critical to guardedly enthusiastic, I retain the title of the original submission, eternalized at Cornell's e-print archive,\cite{QBCAv1} except that now ``critical'' is intended in the sense of Immanuel Kant's critical philosophy.
 
Section \ref{sec.bp}, following this introduction, outlines the Bayesian theory of probability and explains why it is indeed, as QBists claim, the right theory of probability not only for quantum mechanics but for science in general. Section \ref{sec.ca} presents the central affirmations of QBism. According to QBism, quantum theory is an extension or generalization of (Bayesian) probability theory. Section \ref{sec.qa} shows why this is not as obvious as it should be. In a typical axiomatization of the theory, probability tends to be introduced almost as an afterthought, whereas in fact each axiom only makes sense as a feature of a probability calculus.

According to Fuchs,\cite{FuchsPeri} what needs insight is not the quantum-to-classical transition but the classical-to-quantum. Section \ref{sec.cq} offers such an insight. The mathematical apparatus of classical mechanics, too, is a probability calculus. But because the probabilities it serves to assign are trivial, we are free to believe that the probability assigned to a property is 1 not because the property is certain to be found but because the property is actually possessed. Quantum physics differs from classical physics in just one essential respect: it assigns non-trivial probabilities---probabilities between 0 and~1---and this is why probability~1 is \emph{not} sufficient for ``is'' or ``has.'' Arguably the most straightforward way to make room for nontrivial probabilities is to upgrade the mathematical sense of ``state'' from a \hbox{0-dimensional} point (or subset of a phase space) to a 1-dimensional line (or subspace of a Hilbert space). Once this is done, the mathematical apparatus of the theory is a forgone conclusion, provided that one takes for granted the existence of stable objects that ``occupy'' space while being composed of finite numbers of objects that do not ``occupy'' space (which are commonly thought of as pointlike). 

QBists instead hope to eventually be in a position to derive the standard Hilbert space formalism from their formulation of the Born rule, and so to distill the essential characteristic of the quantum world. This fascinating project is outlined in Secs.~\ref{sec.QBBorn} and~\ref{sec.sic}. Section \ref{sec_shifty} brings up one of the most contentious concepts in quantum mechanics, the ``shifty split''\cite{Bell90} between the system under investigation and the means of investigation. For the QBist, the system under investigation is the world experienced by the subject, and the means of investigation is the experiencing subject. Hence there are as many splits as there are users (of quantum mechanics) or agents (in a quantum world), and there is nothing shifty about them. It is, however, possible to ``outsource'' the split. Not only does this \emph{not} infringe the basic tenets of QBism but it actually \emph{frees} QBism from its ambiguities and potential inconsistencies. To set the stage for this maneuver, Sec.~\ref{sec.ovo} introduces the important distinction between objectification---as in ``the problem (or disaster\cite{vF1990}) of objectification''---and objectivation, as in Erwin Schr\"odinger's ``principle of objectivation''.\cite{SchrLifeMindMatter1} The concept of objectivation goes back to Kant, whose theory of science forms the subject of Sec.~\ref{sec.Kant}. 

The main inconsistency of QBism (which initially drove me into near apoplexy) arises from the ambiguity of the term ``external world.'' This can refer to the mind-independently existing reality of the $\Psi$-ontologist, to the QBist's ``world as it is without agents''\cite{FSinQSE} (which, arguably, corresponds to Kant's world-in-itself), or to the QBist's ``common external world'',\cite{FMS2014} which corresponds to the external world as understood by Schr\"odinger and to the subject-matter of physical science as understood by both Kant and Bohr. In Sec.~\ref{sec.outsourcing} the inconsistency is cured by outsourcing the split, i.e., by shifting it from the near boundary of our common external world (in the experiencing subject) to its far boundary (between our common external world and the world as it is without agents). How this shifts affects the QBist way of making sense of EPR-Bohm correlations\cite{EPR,Bohm51} is discussed in Sec.~\ref{sec_nonlocal}.

Today, Bohr is mostly known for his insistence on the necessity of using classical concepts, for attributing this necessity to the need  ``to communicate to others what we have done and what we have learned'',\cite{Bohr-EAPHK,Bohr-APHK} and for the thesis that, out of relation to experimental arrangements, physical quantities are undefined. The links between these requirements, however, belong to a fabric of thought that is no longer widely known. This fabric is the subject of Sec.~\ref{sec.Bohr}. Section \ref{sec_QBB} addresses some QBist misconceptions about Bohr.

The next three sections present in outline an ontology that incorporates the insights of QBism and Bohr.\cite{Mohrhoff_Manifesting,Mohrhoff_NewLight,TWATQM} In Sec.~\ref{sec.beyondQBB} I formulate an interpretive principle that warrants the following conclusion: what presents itself here and now with these properties and what presents itself there and then with those properties is one and the same ``thing.'' In Sec.~\ref{sec.UR} this sole existing ``thing'' is identified with the unspeakable reality of the QBist, which responds to our experimental probings, and the unspeakable reality of Kant, which affects us
in such a way that we have the sensations that we do and are able to construct a shared external world. If we then ask how this unspeakable reality might be related to our common external world, and what light quantum mechanics could throw on their relation, the following answer suggests itself: the quantum laws describe, in the only way it can be described, the process of manifestation, which consists in an (atemporal) transition from the essential unity of UR to the actual {multiplicity} of the manifested world.

Who or what, then, is responsible for the structure of this world? We, because we have constructed it, or UR, because it is or has been manifesting it? The answer lies in the fact that the sole ultimately existing object is also the sole ultimately existing subject. UR manifests the world to us because it manifests it to itself. We are UR just as each particle is UR, and both UR qua ultimate subject and UR qua ultimate object contribute to the manifestation of our common external world. Also explained in Sec.~\ref{sec.pa} is (i)~that the transition takes place in stages, (ii)~that particles, atoms, and molecules, instead of being constituent parts of our common external world, are instrumental in its manifestation, (iii)~why the general theoretical framework of contemporary physics is a probability calculus, and (iv)~why the probabilities are assigned to measurement outcomes. 

Section \ref{sec_now} addresses the hope, expressed by Mermin,\cite{Mermin_Nature} that the QBist conversation can be broadened to provide a solution to the problem of the now. This, however, is a non-starter. Like the color of a Burmese ruby, the experience of \emph{now} is a quality that can only be defined by {ostentation}---by drawing attention to something of which we are aware. It cannot be modeled mathematically.%
\footnote{As was pointed out by Bertrand Russell,\cite{RussellOutline} ``[p]hysics is mathematical, not because we know so much about the physical world, but because we know so little: it is only its mathematical properties that we can discover.''}

Section \ref{sec.sa} provides a summary and a final assessment of the central affirmations of QBism, with additional remarks on the important issue of nonlocality.

\section{Bayesian probability}\label{sec.bp}
{\leftskip\parindent\small Whereas the interpretation of quantum mechanics has only been puzzling us for about 75 years, the interpretation of probability has been doing so for more than 300 years.\par\hfill---Marcus Appleby\cite{Appleby2005}\par}\medskip

\noindent There are, broadly speaking, three ways to understand probability:  (i)~as objective {chance}---a physical {propensity}, {disposition}, or {tendency}; (ii)~as {relative frequency}, and (iii)~as a degree of confidence or belief. While the first two senses are ubiquitous in the quantum-theoretical literature, the third is a rarity%
\footnote{In the first volume of his Caltech lectures,\cite{FLS1} Feynman makes the following comment on a common phraseology, according to which an experimentally determined probability $p$ has an error $\pm\Delta$: ``There is an implication in such an expression that there \emph{is} a `true' or `correct' probability \dots\ and that the observation may be in `error' due to a fluctuation.  There is, however, no way to make such thinking logically consistent.  It is probably better to realize that the probability concept is in a sense subjective, that it is always based on uncertain knowledge, and that its quantitative evaluation is subject to change as we obtain more information.''}
and, in the radical version advocated by QBists, a novelty. 

While the frequentist definition of probability is readily grasped in the abstract, its suffers from a vicious circularity. Frequentists claim that if one repeats a random experiment over and over, independently and under identical conditions, the fraction of trials that result in a given outcome converges to the ``true'' probability of the outcome as the number of trials grows without bound. While they admit that ``anomalous'' frequencies are likely to be obtained as long as the number of trials is finite, they insist that these will be less and less likely. Of course they will be, but not if ``likely'' means ``high relative frequency in the long run.''

Probability is a quantification of \emph{possibility}. If a possibility is something that \emph{may} happen in the actual world, then where is it when it does not happen in the actual world? In a possible world? And where would that be? In someone's mind! How, then, can a probability exist out there, as objective chance? Bayesians therefore insist that probabilities quantify degrees of belief. Returning to probability theory's ancient roots in {gambling} and {betting}, they imagine that coupons are offered for sale, each bearing the words: ``If the event $E$ occurs, the seller of this coupon will pay the buyer one dollar.'' A bettor's assignment of probability $p_B$ to the occurrence of $E$ means that she is willing to pay any amount up to $p_B$ dollars for a coupon, and a seller's assignment of $p_S$ to $E$ means that he is willing to sell a coupon for any amount upwards from $p_S$ dollars.

The lynchpin of Bayesian probability theory is {Bayes' law}, typically written in the form
\be
p(B|A) = \frac{p(A|B)}{p(A|B)\,p(B)+p(A|\overline{B})\,\,p(\overline{B})}
\,p(B)\,,
\label{Bayes}
\ee
where $\overline{B}$ is the negation of $B$ and $p(A|B)$ is the probability of $A$ given~$B$. Bayes' law is a rule for updating personal probability assignments in the light of new information. If your initial probability for $B$ is the prior (probability) $p(B)$, and if a subsequent event~$A$ makes you revise your initial probability assignment, then you should henceforth use the updated (posterior) probability $p(B|A)$. The Bayesian view of probability is currently gaining favor among scientists and philosophers alike. According to Massimo {Pigliucci},\cite{Pigliucci} it is a ``way to think about science and its power and limits that \dots\ beautifully clears up many common misunderstandings that both scientists and the general public seem to have about the nature of science'':
\bq
Bayesian statistical analysis is a good metaphor (some philosophers of science would say a good description) of how science really works. More, it is a good description of how any logical inquiry into the world goes if it is based on a combination of hypotheses and data. The scientist (and in general the rationally thinking person) is always evaluating several hypotheses on the basis of her previous understanding and knowledge on the one hand and on new information gathered by observation or experiment on the other hand.
\eq

\section{The central affirmations of QBism}\label{sec.ca}
QBists take the personalist Bayesian interpretation of probability to a new level. What is confined to the private {experience} of the individual user (of quantum mechanics) or agent (in a quantum world) is not merely the probabilities she assigns to the possible outcomes of a measurement, nor merely the compendium of probabilities we call ``{quantum state},'' but also the data on the basis of which she assigns quantum states, the events to which she assigns probabilities, and the actions she takes. Thus a measurement, QBistically conceived, is any action an agent takes to elicit one of a set of possible experiences, the apparatus being an integral part of the agent, a kind of {prosthetic hand} by which she probes the quantum world, and the actual measurement outcome is the personal experience elicited thereby, returned by the quantum world in response to her probing.\cite{FS2013} And this goes for \emph{any} measurement: ``QBism treats all physical systems in the same way, including atoms, beam splitters, Stern-Gerlach magnets, preparation devices, measurement apparatuses, all the way to living beings and other agents''\cite{FS2015}: ``an action can be anything from running across the street at L'Etoile in Paris (and gambling upon one's life) to a sophisticated quantum information experiment (and gambling on the violation of a Bell inequality)''.\cite{Fuchs_Notwithstanding} The only phenomenon a QBist ``does not model with quantum mechanics is her own direct internal awareness of her own private {experience}''.\cite{FMS2014}

This approach at once disposes of the {conundrum} posed by ``{Wigner's friend}''.\cite{Wigner} In this scenario, Wigner's friend Frida ($F$) has performed a measurement on a system~$S$ with the help of an apparatus~$A$. For Wigner, the state of the combined system $S+A+F$ is believed (by the real Wigner) to change from a coherent superposition of states corresponding to the possible outcomes to a state in which a definite outcome is indicated only when he learns from Frida which outcome she has obtained. For her, the change of $S+A$ takes place earlier, when {she} becomes aware of the outcome. To the real-world Wigner, this appeared to imply that the theory of measurement was logically consistent only ``so long as I maintain my privileged position as ultimate observer''.%
\footnote{As a ``way out of the difficulty,'' Wigner\cite{Wigner} proposed ``that the equations of motion of quantum mechanics cease to be linear, in fact that they are grossly non-linear if conscious beings enter the picture.'' Later he retracted his conclusion that ``consciousness enters the theory unavoidably and unalterably,'' and affirmed instead that consciousness was outside the scope of quantum mechanics.\cite{Wigner1977}}
To QBists, it implies that a quantum state cannot be a user-independent state. For them, a ``true'' quantum state is as much an oxymoron as a user-independent probability.

The probabilities a user assigns to the possible outcomes of a von Neumann measurement are encoded in a density operator~$\hW$, and they are extracted from it via the trace rule $p(\hP)=\hbox{Tr}(\hW\hP)$, to which QBists refer as the \emph{Born rule}. Here $\hP$ stands for the projection operator ``representing'' a possible outcome (for the purpose of extracting its probability from~$\hW$). If a positive-operator-valued measure (or POVM) is used instead of the projector-valued von Neumann measure (or PVM), the Born rule takes the form 
\be
p(\hE)=\hbox{Tr}(\hW\hE),
\ee
where $\hE$ stands for the positive operator ``representing'' (in the above sense) a possible outcome. If a user's prior information about a quantum system is encoded in~$\hW_1$, and if she then obtains the outcome represented by the positive operator~$\hE$, her updated information about the system is given by
\be
	\hW_2=\frac{\hE\hW_1\hE}{\hbox{Tr}(\hE\hW_1\hE)}\,.
\label{QBayes}
\ee
If the outcome is represented by a projector $\hP$, this hold with $\hP$ in place of~$\hE$. Equation (\ref{QBayes}) is the quantum-mechanical form of Bayes' law (Eq.~\ref{Bayes}).

\section{Quantum axiomatics}\label{sec.qa}
{\leftskip\parindent\small If quantum theory is so closely allied with probability theory, why is it not written in a language that starts with probability, rather than a language that ends with it?\par\hfill---Christopher A.\ Fuchs\cite{Fuchs_Perimeter}\par}\medskip

\noindent I cannot but agree with the QBist interpretation of quantum mechanics as a probability calculus, designed to assign probabilities to possible measurement outcomes on the basis of actual measurement outcomes. This fact tends to be obscured by the manner in which quantum mechanics is usually taught. Consider the following fairly typical axiomatization of the theory:

First we are told that the state of a system~$S$ is (or is represented by) a normalized element~$\ket v$ of a Hilbert space~$\cH_S$.

Then we are told that observables are (or are represented by) self-adjoint linear operators on~$\cH_S$, and that the possible outcomes of a measurement of an observable~$\hO$ are the eigenvalues of~$\hO$.

Then comes an axiom concerning the time evolution of states. Between measurements, states evolve according to unitary transformations; at the time of a measurement they evolve as stipulated by the projection postulate. That is, if $\hO$ is measured, the subsequent state of~$S$ is the eigenvector $\ket w$ corresponding to the outcome~$w$, regardless of the previous state of~$S$.

A further axiom stipulates that the states of composite systems are (or are represented by) vectors in the tensor product of the Hilbert spaces of the component systems.

Finally there is an axiom concerning probabilities. If $S$ is (or has been prepared in) the state~$\ket v$, and if we do an experiment to see if $\hO$ has the value $\ket w$, then the probability~$p$ of a positive outcome is $p=|\braket wv|^2$. Furthermore, the expectation value of $\hO$ in the state~$\ket v$ is $\sandwich v\hO v$.

There is much here that is perplexing if not downright wrong. What is meant by saying that the state of a system is (or is represented by) a normalized vector in a Hilbert space? What does it mean to say that observables are (or are represented by) self-adjoint operators? And why is it that the states of composite systems are (or are represented by) vectors in the tensor product of the Hilbert spaces of the component systems? Axioms are supposed to be clear and compelling. The standard axioms of quantum theory lack both desiderata.

There are two possible ways out of this quandary. The first is to try to show that the Born rule is redundant, by deriving it from the remaining axioms. This line of attack, spearheaded by Wojciech Zurek,\cite{Zurek2003a,Zurek2003b,Zurek2005envariance} takes for granted (i)~the existence of events to which probabilities can be assigned, and (ii)~the possibility of decomposing the universe into subsystems. The existence of such events in a deterministic ontology, however, remains mysterious, and so does the possibility of ``carving nature at its joints.'' The other way out of the quandary is to accept the fact that the mathematical apparatus of quantum mechanics is a probability calculus, to show that \textit{every} axiom of the theory owes its physical significance to this fact, and to show that mathematical entities like ``states'' and ``observables'' owe their meanings to the roles they play in assigning probabilities to measurement outcomes on the basis of measurement outcomes.

\section{From classical to quantum}\label{sec.cq}
{\leftskip\parindent\small The thing that needs insight is not the quantum-to-classical transition, but the classical-to-quantum!---Christopher A.\ Fuchs\cite{FuchsPeri}\par}\medskip

\noindent Here goes. The mathematical apparatus of classical mechanics, too, is a probability calculus. But because it is trivial, in the sense that it only assigns the probabilities 0 or~1,  it allows us to think of a state in the classical sense of the word: a collection of possessed properties. We can represent the state of a classical system with $n$ degrees of freedom by a point~$\cP$ in a $2n$-dimensional phase space~$\cS$, and we can represent the positive outcome of an elementary test by a subset~$\cU$ of~$\cS$. The probability of obtaining~$\cU$ is 1 if $\cP\in\cU$, and it is 0 if $\cP\not\in\cU$. Because there are no probabilities that are greater than 0 and less than 1, we are free to believe that the reason why the probability of finding the property represented by~$\cU$ is~1 is that the system \emph{has} this property.

Arguably the most straightforward way to make room for nontrivial probabilities is to upgrade from a 0-dimensional point to a 1-dimensional line.\cite{Mohrhoff-QMexplained} (The virtual inevitability of this upgrade was demonstrated by Josef Jauch.\cite{Jauch}) Instead of representing the probability algorithm or ``state'' associated with a system by a point in a phase space, one represents it by a 1-dimensional subspace of a vector space~$\cV$ equipped with a Hermitian inner product, and instead of representing measurement outcomes by subsets of a phase space, one represents them by (closed) subspaces of~$\cV$. A {1-dimensional} subspace~$\cL$ can be contained in a subspace~$\cU$, it can be orthogonal to~$\cU$, but now there is a third possibility, and this makes room for nontrivial probabilities. $\cL$~assigns probability~1 to outcomes represented by subspaces containing~$\cL$, it assigns probability~0 to outcomes represented by subspaces orthogonal to~$\cL$, and it assigns probabilities greater than~0 and less than~1 to measurement outcomes represented by subspaces that neither contain nor are orthogonal to~$\cL$. The one-to-one correspondence between subspaces and projectors allows us to formulate the following axiom:

\medskip\noindent\textbf{Axiom 1.} {Measurement outcomes are represented by the projectors of a vector space.}

\medskip\noindent What makes the quantum-mechanical probability calculus nontrivial is the existence of incompatible observables. The formal characterization of measurements of compatible observables is also readily established\cite{Mohrhoff-QMexplained}:

\medskip\noindent\textbf{Axiom 2.} {The outcomes of compatible measurements (i.e., measurements of compatible observables) correspond to commuting projectors.}

\medskip\noindent Now imagine two perfect---one hundred percent efficient---detectors $D(A)$ and $D(B)$ monitoring two disjoint intervals $A$ and $B$. If the probabilities $p({A})$ and $p({B})$ are both greater than~0, then it is not certain that $D(A)$ will click, and it is not certain that $D(B)$ will click. Yet if $p({A}\cup {B})=1$, then it \textit{is} certain that either $D(A)$ or~$D(B)$ will click. What makes this certain? The obvious answer is that quantum mechanics is a probability calculus, that the events to which, and on the basis of which, probabilities are assigned, are measurement outcomes, and that measurements are measurements only if they have outcomes.%
\footnote{This also means that quantum mechanics does not give us the probability with which an attempted measurement will succeed, and that it is incapable of formulating sufficient conditions for a successful measurement.}
But now suppose that $A$, $B$, and $C$ are disjoint intervals, and that we have both $p({A})+p({B})+p({C})=1$ and $p({A}\cup{B})+p({C})=1$.  Does it follow that  $p(A\cup B)=p(A)+p(B)$? Because the measurement with three possible outcomes $A$, $B$, $C$ and the measurement with two possible outcomes ${A}\cup{B}$ and ${C}$ require different experimental setups, the answer can only be a postulate, however obvious it may seem:

\medskip\noindent\textbf{Axiom 3.} If $\hA$ and~$\hB$ are orthogonal projectors, then the probability of the outcome represented by the projector $\hA+\hB$ is the sum of the probabilities of the outcomes represented by $\hA$ and~$\hB$, respectively.

\medskip\noindent Armed with Axioms 1--3, one can prove the celebrated theorem by Andrew Gleason,\cite{Gleason} which states that the probability of obtaining the outcome represented by $\hP$ can be extracted from a unique density operator $\hW$ via the Born rule $p(\hP)=\Tr(\hW\hP)$. Gleason's proof holds for vector spaces with at least three dimensions. More recently, the validity of his theorem has been extended to include 2-dimensional vector spaces, by using POVMs instead of PVMs.\cite{Fuchs2001,Busch,Cavesetal}

Probabilities depend on data, which are the (known) outcomes of measurements, and they depend on the time of the measurement to the possible outcomes of which they are assigned. The rule for updating probability assignments on the basis of new data, and the rule for taking account of the time between measurements, are also readily established.\cite{Mohrhoff-QMexplained} In this way we find that the formal apparatus of quantum mechanics can be obtained by upgrading from the 0-dimensional probability calculus of classical mechanics to a probability calculus that represents states by rays in a complex vector space~$\cV$, and that the upgrade is necessitated by the existence of incompatible observables. But whence the need for incompatible observables, and why does $\cV$ have to be complex? These features of the formalism may be seen as consequences of the existence of stable objects that ``occupy'' space while being composed of finite numbers of objects that do not ``occupy'' space (which are commonly thought of as pointlike).\cite{Mohrhoff_Manifesting,Mohrhoff-QMexplained}

\section{QBism and the Born rule}\label{sec.QBBorn}
State vectors and self-adjoint operators, the familiar ingredients of the probability calculus called ``quantum mechanics,'' can be dispensed with, or so QBists claim. The Born rule is fundamental, and in fact it can be expressed entirely in terms of probabilities, as the following will show. 

While a density operator $\hW$ determines a potentially infinite number of probabilities, these cannot all be independent. On a $d$-dimensional Hilbert space, $\hW$ is completely determined by the $d^{\,2}$ probabilities it assigns to the outcomes represented by the positive operators constituting an informationally complete measurement. A measurement on a system with a $d$-dimensional Hilbert space is said to be \emph{informationally complete} if its outcomes are represented by $d^{\,2}$ linearly independent positive operators~$\hE_i$, each proportional to a 1-dimensional projector $\hPi_i$. Any density operator $\hW$ thus corresponds to a vector whose components are the probabilities it associates with the outcomes of an informationally complete measurement, 
\be
p_i=\hbox{Tr}(\hW\hE_i)\,,
\ee
and any POVM $\{\hF_j\}$ corresponds to a matrix whose elements are the conditional probabilities
\be
R_{ji}=\hbox{Tr}(\hPi_i\hF_j)\,.
\ee
The Born rule can therefore be written in the generic form
\be
q_j= f \bigl(\{R_{ji}\},\{p_i\}\bigr)\,,
\label{eq.genericBorn}
\ee
where $f$ depends on the details of the informationally complete measurement $\{\hE_i\}$.

Quantum mechanics thus stands revealed as a generalization of the Bayesian theory of probability. It is a calculus of consistency---a set of criteria for testing coherence between such beliefs as are relevant for rational decision making. As there are no {external} criteria for declaring a probability judgment right or wrong, so there are no {external} criteria for declaring a quantum state assignment right or wrong. The only criterion for the adequacy of a probability judgment or a state assignment is internal coherence between beliefs. The Born rule is not simply a rule for {updating} probabilities, for getting new ones from old. It is a rule for {relating} probability assignments and {constraining} them. It defines ``Bayesian coherence'' for the quantum world. It is \emph{normative}, a rule to guide an agent's behavior in a world that is fundamentally quantum.

There is, however, one significant difference. While the standard rules of probability theory can be demonstrated by arguments showing that a bettor would lose if her bets were not consistent with them, the Born rule is an \emph{empirical} rule. It is a statement about the quantum world, indirectly expressed as a calculus of consistency for bets placed on the outcomes of measurements.

\section{SIC measurements}\label{sec.sic}
The function $f$ in Eq.~(\ref{eq.genericBorn}) takes a particularly intriguing form if the positive operators $\hE_i$ constitute a \emph{symmetric} informationally complete (SIC) measurement. If the positive operators $\hE_i$ constitute a SIC measurement, there exist $d^2$ projectors $\hPi_i=\ketbra ii$ such that $\hE_i=\hPi_i/d$ and $\hbox{Tr}(\hPi_i\hPi_j)=|\braket ij|^2=(d+1)^{-1}$ for $i\neq j$. The density operator can then be given the form\cite{Fuchs_Perimeter,Fuchs2004}
\be
\hW=\sum_{i=1}^{d^2}\left[(d+1)\,p_i-\frac1d\right]\hPi_i=
(d+1)\sum_{i=1}^{d^2}p_i\hPi_i-\hI\,,
\label{eq.bado}
\ee
and the Born rule can be written as
\be
q(\hF_j)=\hbox{Tr}\left(\sum_{i=1}^{d^2}
\left[(d+1)\,p_i-\frac1d\right]\hPi_i\hF_j\right)
=\sum_{i=1}^{d^2}\left[(d+1)\,p_i-\frac1d\right]R_{ji}\,.
\label{eq.bornagain}
\ee
Observe that the right-hand side depends on nothing but marginal and
conditional probabilities. If the positive operators $\hF_j$ are mutually orthogonal projectors $\hP_j=\ketbra jj$, representing the outcomes of a von Neumann measurement, then the corresponding probabilities take the simpler form
\be
q_j=\sum_{i=1}^{d^2}
\left[(d+1)\,p_i-\frac1d\right]\sandwich j{\hP_i}j
=(d+1)\sum_{i=1}^{d^2}\,R_{ji}p_i-1\,.
\label{eq.BornvN}
\ee
The Born probabilities (\ref{eq.bornagain}) and (\ref{eq.BornvN}) pertain to situations in which the SIC measurement $\{\hE_i\}$ is \emph{not} made. If it \emph{is} made, the familiar law of total probability applies, and we have that
\be
q_j=\sum_{i=1}^{d^2}R_{ji}p_i \,.
\label{eq.BornA}
\ee
Contemplate the difference between the right-hand sides of Eqs.~(\ref{eq.BornvN}) and (\ref{eq.BornA}). Clearly, the Born rule is ``nothing but a kind of Quantum Law of Total Probability''.\cite{Fuchs_Perimeter} It contains no hint of the usual formalism of quantum mechanics---just probabilities in and probabilities out.

QBists hope to eventually be in a position to derive the standard Hilbert space formalism from their formulation of the Born rule. And they hope so to distill the essence of quantum mechanics and the essential characteristic of the quantum world. This a fascinating, highly ambitious, and  seriously challenging project. Do SIC measurements even exist? Unfortunately, proofs of their existence are elusive.%
\footnote{This is a deep subject with mathematical tendrils running in many directions.\cite{Applebyetal2015} The construction of SIC operator sets has even been found related to one of the remaining unsolved mathematical problems listed by David Hilbert at the beginning  of the 20th century.\cite{Applebyetal2017}}
As of May 2017, such proofs have been found for all dimensions up to $d{=}151$, and for a few others up to 323.\cite{Fuchs_Notwithstanding} The mood of the QBist community nevertheless is that a SIC measurement should exist for every finite dimension. That said, it must be stressed that the general form of the Born rule, Eq.~(\ref{eq.genericBorn}), does not depend on the existence of SIC measurements; it only presupposes informationally complete POVMs, and these are known to exist for all finite dimensions.

\section{QBism and the ``shifty split''}\label{sec_shifty}
One of the most controversial concepts in the quantum-theoretical literature is  the ``shifty split'' deplored by Bell,\cite{Bell90} otherwise known as Heisenberg cut, between the system under investigation and the means of investigation. For the QBist, the system under investigation is the experienced world, and the means of investigation is the experiencing subject. Accordingly, there are as many splits as there are users or agents, and there is nothing shifty about them. As Mermin explains,\cite{Mermin_shifty}
\bq
Each split is between an object (the world) and a subject (an agent's irreducible awareness of her or his own experience). Setting aside dreams or hallucinations, I, as agent, have no trouble making such a distinction, and I assume that you don't either. Vagueness and ambiguity only arise if one fails to acknowledge that the splits reside not in the objective world, but at the boundaries between that world and the experiences of the various agents who use quantum mechanics.
\eq
Another frequently objected-to concept is the (alleged) irreversibility of the {measurement process}. There is no dearth of arguments to the effect that measurements are irreversible \emph{for all practical purposes}, but there appears to be only one incontrovertible argument why measurements are irreversible \emph{in principle}. This is that outcomes only exist \emph{in experience}, and are therefore endowed with the incontestably irreversible actuality of experience. To the extent that the apparatus is present in {my} experience, it is necessarily classical \emph{for me}. It satisfies the Kantian \emph{principle of thoroughgoing determination},\cite{KantCPR1} according to which the following must hold for each thing at all times: of any two possible predicates that are opposites of each other, one must apply to it. In other words, in my experience there are no coherent superpositions of {outcome-indicating} properties. Only if the apparatus is not present in my experience, can I treat it as a quantum object rather than as a measurement apparatus, and in this case QBism requires that I treat it so.

\section{Objectification versus objectivation}\label{sec.ovo}
Most current literature on the ``big'' measurement problem\cite{Pitowsky2006}---the problem of explaining how measurement outcomes come about dynamically, a.k.a.\ the problem (or ``disaster''\cite{vF1990}) of objectification---still follows the first rigorous formulations of the problem in the monographs of von Neumann\cite{vN} and Pauli.\cite{PauliGP} In this literature a measurement is modeled as a two-stage process: the so-called {premeasurement} ($pm$), which takes the prepared state of the system ($S$) and the neutral state of the apparatus ($A$) to a bi-orthogonal entangled state of the combined system ($S{+}A$), and the coming into existence of an outcome called \emph{objectification} ($ob$):
\be
\ket{A_0}\ket\psi
                 \;\stackrel{(pm)}{\longrightarrow}\;
\sum_{k}c_k\ket{A_k}\ket{q_k}
                  \;\stackrel{(ob)}{\longrightarrow}\;
\ket{A(q)}\ket{q}.
\label{eq_pmob}
\ee
We are made to understand that, initially, the apparatus is \emph{in} the neutral state $\ket{A_0}$ and the system is \emph{in} the prepared state $\ket\psi$, and that, in the end, the measured observable $Q$ has the value~$q$, and the apparatus indicates that this is the case. Except that the final state is never reached; this is the gist of \emph{insolubility theorems} for the objectification problem proved by Peter Mittelstaedt\cite{Mittelstaedt98} and by Busch, Lahti, and Mittelstaedt.\cite{BLM96} 

While objectification is hoped (against hope) to take us from a realistically construed quantum state to what is experienced by us, objectivation begins with what is experienced by us. Unlike objectification, objectivation is not a problem; it is a principle that was extensively discussed by Schr\"odinger,\cite{SchrLifeMindMatter1} who according to Fuchs \emph{et al.}\cite{FMS2014} took ``a QBist view'' of science. In an essay written during the last year of his life, Schr\"odinger\cite{SchrWhatIsReal} expressed his astonishment that ``despite the absolute hermetic separation of my sphere of consciousness'' from everyone else's, there is ``a far-reaching structural similarity between certain parts of our experiences, the parts which we call external; it can be expressed in the brief statement that we all live in the same world.'' This far-reaching structural similarity, he avowed, ``is not rationally comprehensible. In order to grasp it we are reduced to two {irrational, mystical hypotheses},'' one of which is ``the so-called {hypothesis of the real external world}''.%
\footnote{The other irrational, mystical hypothesis, which he endorsed, was that ``we are all really only various aspects of the One''.\cite{SchrWhatIsReal} The multiplicity of ``minds or consciousnesses \dots\ is only apparent, in truth there is only one mind. This is the doctrine of the {Upanishads}. And not only of the Upanishads\dots. I should say: the over-all number of minds is just one. I venture to call it indestructible since it has a peculiar timetable, namely mind is always \emph{now}''.\cite{SchrLifeMindMatter2} The last sentence has bearing on Sec.~\ref{sec_now}.}
Schr\"odinger left no room for uncertainty about what he thought of this hypothesis:
\bq
I get to know the external world through my sense-perceptions. It is only through them that such knowledge flows into me; they are the very material out of which I construct it. The same applies to everyone else. The worlds thus produced are, if we allow for differences in perspective, etc., very much the same, so that in general we use the singular: world. But because each person's sense-world is strictly private and not directly accessible to anyone else, this agreement is strange; what is especially strange is how it is established. Many people prefer to ignore or gloss over the strangeness of it, explaining the agreement by the existence of a real world of bodies which are the causes of sense-impressions and produce roughly the same impression on everybody. But this is not to give an explanation at all; it is simply to state the matter in different words.  In fact, it means laying a completely useless burden on the understanding. 
\eq
It means translating the statement ``everybody agrees about something'' into the  statement ``there exists a real world which causes everybody's agreement.'' Instead of explaining the fact expressed by the first statement, the second renders it even more inexplicable, for if a causal relation exists between this so-called real world and those aspects of our experiences about which there is agreement, it is something we cannot know. The causal relations we can know belong to the common part of our internal, subjective experiences---the part which is amenable to {objectivation}, to being represented as a world about which we can think, and in which we can act, \emph{as if} it existed by itself, independent of our thoughts and perceptions (though in fact it does not). This idea goes back to Immanuel Kant, a central figure in modern philosophy, who was a major philosophical source for Schr\"odinger, and who therefore deserves the following philosophical excursion.

\section{The world according to Kant}\label{sec.Kant}
{\leftskip\parindent\small In my opinion, those who really want to understand contemporary physics---i.e., not only to apply physics in practice but also to make it transparent---will find it useful, even indispensable at a certain stage, to think through Kant's theory of science.\par\hfill---Carl Friedrich von Weizs\"acker\cite{vW}\par}\medskip

\noindent Kant owes his fame in large part to his successful navigation between the Scylla of commonsense {realism} and the Charybdis of {idealism}. What allowed Kant to steer clear of both horns of the dilemma, was a dramatic change of strategy. Instead of trying to formulate a metaphysical picture of the world consistent with Newtonian mechanics, as he had done during the pre-critical period of his philosophy, he inquired into the conditions which the ``{manifold of appearances}'' had to satisfy in order to have objectivizable content%
\footnote{I use \emph{objectivize} and \emph{objectivizable}, rather than the more common words \emph{objectify} and \emph{objectifiable}, to emphasize their specific Kantian connotations and to alert the reader to the difference between the concepts of objectivation and objectification.}%
---the conditions, in other words, that make \emph{Erfahrung} ({experience} in the sense of {empirical knowledge}) possible. The fact that we have empirical knowledge then proves that these conditions are satisfied. And the fact that these conditions are satisfied implies the existence of universal and necessary laws of nature, the most important of them being the {law of causation}, according to which ``[a]ll alterations occur in accordance with the law of the connection of cause and effect''.\cite{KantCPR2}

We construct empirical knowledge with the help of logical and spatiotemporal relations. The logical relation between subject and predicate allows us to think of perceptions as being connected according to the {category of substance}; we are able to think of them as properties of a substance, connected to it as predicates are connected to a subject.%
\footnote{Because Kant associated substance with mass, which allowed him to formulate a conservation law for the ``quantum of substance'',\cite{KantCPR3} he did not consider individuality constitutive of the concept of substance. For him, the objectivation of individuality was a matter of establishing transtemporal identity, and this called for a law of causal evolution.}
The logical relation between antecedent and consequent (if \dots\ then\dots) allows us to think of perceptions as being connected according to the {category of causation}; we are able to think of properties perceived at different times as causally connected. And the disjunctive logical relation (either\dots\ or\dots) allows us to think of perceptions as connected according to the category of community or reciprocity; we are able to think of properties perceived in different locations as connected through a reciprocal action between the substances to which they belong.

Moreover, to be able to think of perceptions as properties of substances, or as causally connected, or as affecting each other, the connections must be regular. For perceptions to be perceptions of a particular kind of thing (say, an elephant), they must be connected in an orderly way, according to a concept denoting a lawful concurrence of perceptions. For perceptions to be causally connected, like (say) lightning and thunder, they must fall under a causal law, according to which one perception necessitates the subsequent occurrence of another. And for perceptions to be reciprocally connected, like (say) the Earth and the Moon, they must affect each other according to a reciprocal law, such as Newton's law of gravity. It is only through lawful connections in the {manifold of appearances} that we are able to think of appearances as perceptions of a self-existent system of re-identifiable interacting objects, from which the experiencing subject can remove itself. 

Time and space, according to Kant, are the pure forms of \emph{Anschauung}---a hard-to-translate German word, which covers both imagination and sensory perception. Objects imagined and objects perceived share the same space. There is not a real space that contains real things over and above a subjective space that contains perceptions of real things. Space and time are aspects of the world because the world is constructed by us in these pure forms of the human sensorium. By establishing a causal relation, we at the same time establish an objective temporal relation, and by establishing a reciprocal relation, we at the same time establish an objective spatial relation as well as an objective relation of simultaneity.%
\footnote{While it is consistent with Kant's epistemology to think of the reciprocal action between two objects as grounding both kinds of relation, Kant himself saw the objectivity of spatial relations grounded in the objectivity of motion in space.}

Kant's conception of space and time solved a problem that had troubled many of his contemporaries: how can geometry, whose laws apply to objects we construct in the subjective space of our imagination, also apply to the physical world? Geometry applies to the physical world because the space in which we construct the objects of geometry is also the space in which we construct the physical world. It is noteworthy that Kant's argument  applies to geometry in general, and thus to whichever geometry is best suited to formulating the laws of physics, not to Euclidean geometry specifically, though this was the only geometry known in Kant's time. It applies to the spacetime geometry of the special theory of relativity, in which the invariant velocity is finite rather than infinite and simultaneity is relative rather than absolute. And it applies to the pseudo-Riemannian geometry of the general theory of relativity, in which the straight lines traced by freely moving objects in the mechanics of Newton are replaced by the geodesics traced by light rays and freely falling objects. It may even be said that Kant's conception of the relationship between geometry and physics set in motion a series of re-conceptualizations of this relationship, and that this eventuated in Einstein's theories of relativity.\cite{Friedman2009}

Kant's reading of Newtonian mechanics also dispelled many qualms that had been shared by thinkers at the end of the 18th Century---qualms about the purely mathematical nature of Newton's theory, about the unintelligibility of instantaneous action at a distance, and about Galileo's principle of relativity. The mathematical nature of Newtonian mechanics was justified, not by the Neo-Platonic belief that the book of nature was written in mathematical language, but by being a precondition of scientific knowledge. It is only the mathematical regularities among our perceptions that permit us to conceive of the latter as perceptions of an objective world. Newton's refusal to explain gravitational action at a distance was similarly justified by the fact that the only intelligible causality available to us consists in regular mathematical relations between perceptions. For the Moon to be causally related to the Earth was for the Moon to stand in a regular mathematical relation to the Earth. 

As for the Galilean principle of relativity, according to which motion is always relative, it was a direct consequence of the fact that regular mathematical relations only exist between different perceptions, and thus only between different events or different objects, but never between a particular perception and either space or time itself. Here, too, it would be an anachronism to argue that Kant singled out Galilean relativity; while the latter was the only kind of relativity known in his time, his argument holds for whichever principle of relativity is best suited---or most convenient%
\footnote{``One geometry cannot be more true than another, it can only be more convenient.''---Henri Poincar\'e.\cite{Poincare1905}}%
---to formulating the laws of physics.

\section{Outsourcing the ``shifty split''}\label{sec.outsourcing}
More often than not, what QBists have in mind when they speak of the ``{external world}'' is the world ``we have all negotiated with each other''.\cite{MerminQBnotCop} As Fuchs \emph{et al.}\cite{FMS2014} explain, this world 
\bq
is constrained by the fact that different agents can communicate their experience to each other, limited only by the extent that personal experience can be expressed in ordinary language. Bob's verbal representation of his own experience can enter Alice's, and vice-versa. In this way a common body of reality can be constructed, limited only by the inability of language to represent the full flavor---the ``qualia''---of personal experience.
\eq
It is this common body of constructed reality that the na\"\i ve realists reify, thereby ``removing from the story any reference to the origins of our common world in the private experiences we try to share with each other through language''.\cite{MerminQBnotCop} This common world is none other than the objective world as understood by Schr\"odinger, the part of our experience which is amenable to {objectivation}, about which we can think as if it existed independently of our thoughts and perceptions (though in reality it does not).%
\footnote{Schr\"odinger, too, emphasized the role that {language} plays in establishing the (to him) astonishing correspondence between the content of one sphere of consciousness and that of another\cite{SchrWhatIsReal}: ``What does establish it is \emph{language}, including everything in the way of expression, gesture, taking hold of another person, pointing with one's finger and so forth, though none of this breaks through that inexorable, absolute division between spheres of consciousness.''}
But when Mermin goes on to explain that science is ``a collaborative human effort to find, through our individual actions \emph{on the world} and our verbal communications with each other, a model for what is common to all of our privately constructed external worlds'' (emphasis added),\cite{MerminQBnotCop} the word ``world'' assumes a second meaning. It no longer refers only to our private external worlds but also to the world on which we act. Here one must bear in mind that QBists take the measurement apparatus to be ``an extension of the agent himself, \dots like a prosthetic hand''\cite{Fuchs_Notwithstanding}:
\bq
The agent, through the process of quantum measurement stimulates the world external to himself. The world, in return, stimulates a response in the agent that is quantified by a change in his beliefs---i.e., by a change from a prior to a posterior {quantum state}. Somewhere in the structure of those belief changes lies quantum theory's most direct statement about what we believe of the world as it is without agents.\cite{FSinQSE}
\eq
The world the agent touches with his prosthetic hand---the world on which he acts, which he stimulates, and which responds---is not our common body of constructed reality, for the latter contains his prosthetic hand. It is the unspeakable world ``as it is without agents,'' which only becomes speakable through the manner in which it is stimulated (i.e., by saying in ordinary language what the agent has done) and through the manner in which it responds (i.e., by saying in ordinary language what the agent has learned). But this means that there is another boundary where the Heisenberg cut can be placed: between ``what is common to all of our privately constructed external worlds'' and ``the world as it is without agents,'' which is stimulated and which responds. As there is a ``near'' boundary to our common external world, situated between it and the private experiences in which it originates, so there is a ``far'' boundary to our common external world.

How does shifting the shifty split to the far boundary of our common external world affect the QBist stance? Where Wigner and his friend are concerned, it remains true that they assign different, agent-specific quantum states. But if an agent has firm reasons to believe that an outcome is present in another agent's experience, he has equally firm reasons to believe that an outcome exists in our common external world. In this case, it makes no sense for Wigner to assign to Frida and her system a coherent superposition; he is perfectly justified (not just ``for all practical purposes'') to assign a mixture representing his ignorance of what Frida has found. Much the same applies to the QBist way of making sense of EPR-Bohm correlations,\cite{EPR,Bohm51} to be discussed next.

\section{QBism and nonlocality}\label{sec_nonlocal}
According to the usual, third-person account, the outcomes of measurements made on two EPR-correlated systems come into being at two spacelike separated sites. This seems to imply a ``spooky'' (i.e., nonlocal and superluminal) connection. According to the QBist first-person account, Alice's outcome comes into being in \emph{her} {experience}, while Bob's outcome comes into being in \emph{his} experience. Insofar as the experiences of a given agent can be be located, they are located where the agent is located, i.e., in our common external world. They can then only be separated in time. QBists have therefore claimed that quantum mechanics is ``\emph{explicitly} local''.\cite{FMS2014} But does it make sense to locate experiences? To find out, let us turn once more to Schr\"odinger. 

As Schr\"odinger saw clearly, by objectivizing certain aspects of our experiences, treating them as aspects of a shared {external world}, we (experiencing and thinking subjects) \emph{exclude} ourselves from this world: ``We step with our own person back into the part of an onlooker who does not belong to the world''.\cite{SchrLifeMindMatter3} In other words, ``[m]ind has erected the objective outside world \dots\ by the simplifying device of excluding itself---withdrawing from its conceptual creation.'' If we then mistake the mind's creation for a self-existent real world, we are left with no choice but to re-insert the mind: ``I so to speak put my own sentient self (which had constructed this world as a mental product) back into it---with the {pandemonium} of disastrous logical consequences'' that flow from this error, such as ``our fruitless quest for the place where mind acts on matter or vice-versa.'' The upshot is that locating experiences in our common external world (let alone anywhere else) is not an option.

The claim that quantum mechanics is ``\emph{explicitly} local in the QBist inter\-pretation''\cite{FMS2014} is defensible if it means that nothing happens to Bob's system as a result of Alice's performing her measurement or becoming aware of its outcome. The outcome exists in her mind, and she can use it to assign probabilities to possible future experiences, such as receiving from Bob a report of his outcome. It is, however, necessary to distinguish between correlations in the experience of one and the same person (which \emph{can} exist between spacelike separated events in a user's past light cone) and correlations between events in no one's experience. The QBist interpretation does not proffer a ``{view from nowhere}''.\cite{Nagel_Nowhere} But if ``QBism don't do third-person''\cite{Fuchs_Notwithstanding} then it cannot be the case that QBism ``gives each {quantum state} a home \dots\ localized in space and time---namely, the physical site of the agent who assigns it''.\cite{Fuchs_Perimeter} A quantum state has its home in the agent's mind, not at any physical site (in our common external world) let alone in no one's experience (in which there are no physical sites). Instead of asserting that QBism is explicitly local, QBists therefore ought to assert that QBism is neither local nor nonlocal in any realist sense of these terms.

The additional claim that QBism ``expels once and for all the fear that quantum mechanics leads to `spooky action at a distance'\,''\cite{Fuchs_Perimeter} is a claim about EPR-Bohm correlations in a mind-independently existing external world. But physics does not deal with such a world. Something fearsome is implied by the existence of EPR-Bohm correlations only if one forgets that physics deals with our common external world. Then one worries how, in the absence of a common cause, measurement outcome in spacelike relation can be so spookily correlated. If one keeps in mind that that measurement outcomes are responses from a ``domain'' or ``world'' that is not subject to the pure forms of human experience, one realizes that the answer to this question is beyond the reach of the concepts available to us, based as they are on the pure form of human experience.%
\footnote{Kinematical concepts in particular owe their meanings to the spatiotemporal structure of human experience. For position and time this is obvious enough but it also holds for energy, momentum, and angular momentum, related as they are to the homogeneity of time and the homogeneity and isotropy of space.}

\section{Niels Bohr}\label{sec.Bohr}
{\leftskip\parindent\small It is often said that a work of genius resists categorization. If so, Bohr's philosophical viewpoint easily passes this criterion of greatness. Surely this is one of the reasons for the commonplace complaints over Bohr's ``obscurity.''---Henry J.\ Folse\cite{Folse}\par}\medskip

\noindent Many of Bohr's contemporaries saw it as their task to clarify Bohr's views, which is how they came to be appropriated by different philosophical schools. By the 1960s, many versions of Bohr's views were in circulation, some diametrically opposed to one another. It therefore comes as no surprise that QBists distance themselves not only from $\Psi$-ontology but also from views that came to be attributed to Bohr. I feel, on the contrary, that QBism and Bohr complement each other, and this not in Bohr's various senses of complementarity, which imply mutual exclusion. QBism not only has the potential to lead the physics community (back) to a proper understanding of Bohr but also, in doing so, to a real understanding of quantum mechanics itself. Arguably, therefore, another clarification of Bohr's views---at any rate, another inference to the best explanation of what Bohr meant to say---is warranted.

Today, Bohr is mostly known for his insistence on the necessity of using classical concepts, for attributing this necessity to the need  ``to communicate to others what we have done and what we have learned'',\cite{Bohr-EAPHK,Bohr-APHK} and for the thesis that, out of relation to experimental arrangements, physical quantities are undefined: ``the procedure of measurement has an essential influence on the conditions on which the very definition of the physical quantities in question rests''.\cite{BohrQMPR} The links between these requirements, however, belong to a fabric of thought that is no longer widely known. 

The thesis that physical quantities are defined by the (classically described) experimental conditions in which they are measured, has been spectacularly borne out by the {no-go theorems} of Bell,\cite{Bell64} Kochen and Specker,\cite{KochenSpecker} and Klyachko \textit{et al.}\cite{Klyaetal2008}.%
\footnote{Bohr's thesis of the indispensability of classical language has also been vindicated by more recent developments in particle physics. As Brigitte Falkenburg concludes in her monograph \emph{Particle Metaphysics},\cite{Falk2007} this thesis ``has remained valid up to the present day\dots. Indeed, the use of the familiar physical quantities of length, time, mass, and momentum--energy at a subatomic scale is due to an extrapolation of the language of classical physics to the non-classical domain.''}
What were Bohr's reasons for insisting on it? Like Kant, Bohr came to regard space and time as the warp and woof of human experience, to which he frequently referred as ``our forms of perception.'' The new fundamental constant introduced by Planck, however, implied the existence of limits to the \emph{objectivation} of our forms of perception. And since physical concepts owe their meanings to these forms, it also implied limits to the applicability of these concepts. Yet ``it would be a misconception to believe that the difficulties of the atomic theory may be evaded by eventually replacing the concepts of classical physics by new conceptual forms''.\cite{BohrATDN}

Indeed, physical concepts are physical (rather than purely mathematical) because something in human experience corresponds to them. But the only concepts corresponding to aspects of human experience are the concepts of classical physics. Only classical concepts, therefore, can perform the crucial task of mediating between mathematics and experience. Hence, ``it continues to be the application of these concepts alone that makes it possible to relate the symbolism of the quantum theory to the data of experience''.\cite{BohrATDN} A detector is needed not only to indicate the presence of something somewhere but also---and in the first place---to give ``somewhere'' a meaning and thus to make it possible to attribute to something the property of being somewhere. Speaking more generally, a measurement apparatus is needed not only to indicate the possession of a property by a quantum system but also---and in the first place---to realize a set of properties so as to make them available for attribution to the system.

It is of course tempting to think of the {measurement apparatus} as a sort of extension of the observing subject, as QBists do. Heisenberg,\cite{Heisenberg1935} too, once drew a dividing line between ``the apparatus which we \dots, in a way, treat as part of ourselves'' and the systems investigated. Pauli,\cite{Pauli1955} in a letter to Bohr, went further: 
\bq
As it is allowed to consider the instruments of observation as a kind of prolongation of the sense organs of the {observer}, I consider the impredictable change of the state by a single observation \dots\ to be \emph{an abandonment of the idea of the isolation (detachment) of the observer from the course of physical events outside himself}. [original emphasis]
\eq
Bohr would have none of this. If instruments form part of the organs of observation, then the observed systems \emph{themselves} are not given in the spatiotemporal conditions of human experience, and the concepts which derive their meanings from these conditions cannot be applied to them. What is not directly accessible to our senses becomes thinkable only through such aspects of instruments as are directly accessible to our senses. These aspects---for instance, the regions monitored by an array of detectors---\emph{define} the properties that can be attributed to a quantum system. What cannot be separated from the object of investigation is not the subject, which remains the same detached observer it had been before quantum physics came along, but the means of investigation. The business of physics, consisting in ``the development of methods for ordering and surveying human {experience},'' therefore remains ``objective in the sense that it can be unambiguously communicated in the common human language''.\cite{BohrE58-62}

There are two aspects to achieving objectivity. The main concern for Kant was the possibility of empirical knowledge, which required the synthesis of appearances into a system of \emph{objects}. For him the business of our cognitive faculty was to ``work up the raw material of sensible impressions into a cognition of objects''.\cite{KantCPR5} On the face of it, the main concern for Bohr was the ability ``to communicate to others what we have done and what we have learned.'' Both aspects imply each other. Organizing appearances into a system of objects involves a shared repertoire of concepts. It is a \emph{shared} repertoire because the concepts involved owe their meanings to \emph{our} forms of perception and to the logical or grammatical structure of \emph{our} thought or language. And the possibility of saying what we have done and what we have learned presupposes the familiar object-like organization of the flux of appearances.

Kant never considered the possibility that experimental activity may play the constitutive roles he ascribed to our cognitive faculties of {intuition} (\emph{Anschauung}) and thought. The insight that property attributions are \emph{contextual}, in the sense that they cannot be separated from specific experimental arrangements, is Bohr's. The object-like organization of experience, which relies on (among other things) the possibility of re-identifying bearers of properties through time, cannot be extended into the atomic domain. It stops at the apparatus, at the experimental situation, and so does the reach of our language. Once the attributable physical properties are contextualized, because their defining contexts are no longer compatible, only probabilistic predictions remain feasible---correlations between what we have done and what we have learned, which constitute ``a purely symbolic scheme permitting only predictions \dots\ as to results obtainable under conditions specified by means of classical concepts''.\cite{Bohr-APHKa}

What is understood nowadays by ``interpreting quantum mechanics'' is to take the mathematical formalism as the foundation, and to then figure out how it relates to what happens in the laboratory, in the macroscopic world, or in our experience. For Bohr, experience was the foundation: ``in our description of nature the purpose is not to disclose the real essence of the phenomena but only to track down, so far as it is possible, relations between the manifold aspects of our experience''.\cite{BohrATDNa} (The plural is used advisedly, as the following section will show.) But when, in the 1950s, David Bohm\cite{BohmHV} presented a hidden-variables interpretation and Hugh Everett III\cite{EverettRSF} put forward his relative-state interpretation, Heisenberg,\cite{Heisenberg_PP} entering the fray, argued that the Copenhagen interpretation was the only viable interpretation, thereby transforming Bohr's views into just another interpretation of a mathematically formulated theory. 

Historically, Bohr's reply\cite{BohrEPRreply} to the argument by Einstein, Podolsky, and Rosen\cite{EPR} was taken as a definitive refutation by the physics community. By the time interpreting quantum mechanics had become a growth industry, Bohr's perspective was lost. His paper, which only treated the mathematical formalism in a footnote, was seen as missing the point. Considered on a par with a proliferating multitude of interpretations, his epistemological reflections came to be seen as amateurish, ad hoc, or worse.

In the course of the nineteenth century, most philosophers and apparently many of the scientists who cared to reflect on the nature of science came to  adopt (or, in some cases, reinvent) Kant's epistemology. It alleviated the loss of na\"\i ve realism by making it possible to think and behave \emph{as if} na\"\i ve realism were true. The advent of quantum mechanics dealt a severe blow to this comfortable attitude. Today's futile attempts to beat ontological sense into an irreducibly nontrivial probability calculus are born of the hope to return to that state of innocence. Quantum mechanics practically compels us to adopt the idea that was at the very core of Kant's theory of knowledge and constitutes its truly original contribution to philosophy, namely, that things and events, far from being elements of a mind-independent reality, are objectivizable elements of our experience.

Quantum theory reminds us of the roles perception and conception play in the construction and constitution of the objective world, and no one has seen this as clearly as Bohr. Bohr departs from and goes beyond Kant in only one essential respect---his realization that the {objectivation} of the spatiotemporal conditions of human experience is limited. The behavior of quantum systems therefore is consistent neither with Kant's principle of thoroughgoing determination nor with the universal validity of Kant's \emph{a priori} laws. The latter are preconditions for the possibility or organizing experience into objects, and this stops at the apparatus. What takes its place is the contextuality of property-attributions and their complementarity, which offers ``a natural generalization of the classical mode of description''.\cite{BohrQPRDAP} It is ``the mutual exclusion of any two experimental procedures, permitting the unambiguous definition of complementary physical quantities, which provides room for new physical laws'.\cite{Bohr-APHKb} It therefore is an irony that Bohr, seeing Kant as arguing for the universal validity of {classical concepts}, regarded {complementarity} as an alternative to (rather than as an extension or generalization of) Kant's theory of science, thus drawing the battle lines in a way which put {Kant} and himself on opposing sides.

\section{QBism and Bohr}\label{sec_QBB}
It is another irony that QBists tend to draw their battle lines in a way which puts Bohr and themselves on opposing sides, this notwithstanding the fact that ``QBism agrees with Bohr that the primitive concept of \emph{experience} is fundamental to an understanding of science''.\cite{FMS2014} The reason for this appears to be at least in part failure to distinguish---at any rate, distinguish \emph{consistently}---between our common body of constructed reality and the world ``as it is without agents.'' Consider, for instance, this statement by Fuchs \emph{et al.}\cite{FMS2014}:
\bq 
when we attempted to understand phenomena at scales not directly accessible to our senses, our ingrained practice of divorcing the objects of our investigations from the subjective experiences they induce in us got us into trouble.
\eq
By identifying the objects of our investigations with what induces subjective experiences in us, Fuchs \emph{et al.} appear to subscribe to the na\"\i ve realism implicit in the standard scientific account of sensory perception, which is demonstrably false. On this account, external objects emit photons and/or sound waves, which stimulate peripheral nerve endings, which send signals to the brain, where neural processes miraculously give rise to perceptual experience. But then it follows that there is no way we could have access to the aforesaid external objects; we only have access to perceptual experience.%
\footnote{This has been known at least since the Greek philosopher-poet {Xenophanes} pointed out, some twenty-five centuries ago, that even if our minds represented the world exactly as it was, we could never know that this was the case.}
What Fuchs \emph{et al.} actually mean, of course, is that our interpretive woes are caused by our divorcing the objects of our investigations from our subjective experiences---period.

Failing to distinguish between the two kinds of external world---the one we have objectivized and the one which exists in no one's experience---Fuchs \emph{et al.} attribute to Bohr the ``view that measurement outcomes belong to an objective (`classical') domain that is independent of agents and/or their experience.'' This has never been Bohr's view. For Bohr, a domain enjoying this kind of independence either did not exist or did not fall under the purview of physics.%
\footnote{Petersen\cite{Petersen1963} has Bohr saying that ``[i]t is wrong to think that the task of physics is to find out how nature is. Physics concerns what we can say about nature.''}
When Bohr wrote that the description of atomic phenomena has ``a perfectly objective character,'' it was ``in the sense that no explicit reference is made to any \emph{individual} observer'' (emphasis added),\cite{BohrE58-62} not in the sense that no explicit reference was made to observers. The reason empirical knowledge is objective, and therefore knowledge (and therefore objective), is that it is communicable. It is communicable because the concepts used by us are grounded in the forms of perception common to us and in the grammatical structure of a common language.

Mermin thinks that what Bohr meant by ``experience'' was ``almost certainly \dots\ the objective readings of large classical instruments and not the personal experience of a particular user of quantum mechanics''.\cite{MerminQBnotCop} Objective readings of large classical instruments and the personal experience of a particular user are not the only possible meanings of ``experience,'' as Mermin is well aware, writing as he does that  ``[s]cience is a collaborative human effort to find \dots\ a model for what is common to all of our privately constructed external worlds''.\cite{MerminQBnotCop} The concept of science enunciated here is close to Kant's concept of ``experience''---\emph{Erfahrung}, empirical knowledge---which is close to Bohr's. Sadly, QBists consistently fail to realize how close their views are to Bohr's, and how far Bohr's views are from instrumentalism, of which both they and Bohr are accused by $\Psi$-ontologists.

As previously indicated, a measurement in QBism is ``any action an agent takes to elicit a set of possible experiences,'' and ``[t]he measurement outcome is the particular experience of that agent elicited in this way''.\cite{FMS2014} But if measurement outcomes are located in experience, then so are instruments and so is the rest of the classical world. The classical world (including the readings of large classical instruments) is objective because we have constructed it by a collaborative human effort. This construction, this collaborative human effort presupposes the existence of a common language, a shared repertoire of concepts, which is itself the product of a collaborative human effort. Indeed, the creation of a common language cannot be separated from the construction of a {common} external world containing the referents of the common language. The two go hand in hand.

When Mermin claims that ``[o]rdinary language comes into the QBist story in a more crucial way than it comes into the story told by Bohr,'' his contention is that language merely serves to compare ``our privately constructed external worlds''.\cite{MerminQBnotCop} ``Language is the only means by which different users of quantum mechanics can attempt to compare their own private experiences.''  In point of fact, there is no such thing as a privately constructed external world; nobody can privately construct an external world. Language does not serve to compare private worlds; language is the tool by which we have constructed a shared external world. Nobody can even have experiences---experiences that can be articulated, experiences amounting to empirical knowledge---in the absence of a common world to which the concepts of a common language refer. Arguably, therefore, ordinary language comes into the story told by Bohr in a more crucial way than it comes into the QBist story.

Again, contrasting QBism with what he takes to be Bohr's view, Mermin writes that ``measurement outcomes in QBism are necessarily classical, in a way that has nothing to do with language.'' Bohr, we are given to understand, attributed the classical character of measurement outcomes to language. In actual fact, he attributed the classical character of measurement outcomes to the ``irreversibility implied in the very concept of observation'',\cite{BohrE58-62p92} and by this he meant the incontestable irreversibility of subjective experience, not the irreversibility of physical or chemical processes leading to permanent marks in no one's experience. If physics only permits predictions ``as to results obtainable under conditions specified by means of classical concepts'',\cite{Bohr-APHKa} it is not only because the latter are needed to link the ``purely symbolic scheme'' to human experience but also because human experience alone can provide the irreversible definiteness without which there would be no obtainable results. Bohr, it would seem, took a fundamentally QBist view.

The QBists' incomprehension of Bohr's views goes so far as to blame him (along with his student Heisenberg) for the measurement problem: 
\bq
The Founders of quantum mechanics were already aware that there was a problem. Bohr and Heisenberg dealt with it by emphasizing the inseparability of the phenomena from the instruments we devised to investigate them. Instruments are the Copenhagen surrogate for experience. Being objective and independent of the agent using them, instruments miss the central point of QBism, giving rise to the notorious measurement problem, which has vexed physicists to this day. \cite{FMS2014}
\eq
The aforementioned ``big'' measurement problem is the creation of von Neumann.\cite{vN} For Bohr, a measurement was a holistic phenomenon, which cannot be dissected into a unitary evolution and a subsequent collapse or objectification. Nor did the ``small'' measurement problem\cite{Pitowsky2006}---the question why we do not observe superpositions of macroscopically distinct situations---arise for Bohr. Nor did the question of what makes a situation macroscopic.  Observed situations are either identical or distinct. Observing a superposition of distinct situations is therefore a contradiction in terms.

\section{Beyond QBism and Bohr}\label{sec.beyondQBB}
{\leftskip\parindent\small Atoms---our modern atoms, the ultimate particles---must no longer be regarded as identifiable individuals. This is a stronger deviation from the original idea of an atom than anybody had ever contemplated. We must be prepared for anything.\par\hfill---Erwin Schr\"odinger\cite{SchrNGSH}\par}\medskip

\noindent Quantum systems straddle the boundary between our common external world and the unexperienced ``world'' on the other side. They belong to the former insofar as their properties depend on experimental contexts, and they belong to the latter insofar as they respond to our experimental probings. QBists aim to distill the essential characteristic of the quantum world from the Born rule. For Bohr there was no quantum world\cite{Petersen1963}: ``the physical content of quantum mechanics [was] exhausted by its power to formulate statistical laws governing observations obtained under conditions specified in plain language''.\cite{Bohr-EAPHKp12} (Once again: by ``observations'' Bohr meant experiences, and the necessity of specifying the experimental conditions in plain language arose from the need to say something that made physical rather than merely mathematical or merely qualitative sense.)

Bohr would have insisted (if he did not actually say it) that a measurement outcome experienced by even a single user/agent was to be regarded as a fact belonging to our common external world. Situating the correlata of the quantum-mechanical correlation laws in our common external world---i.e., neither in a mind-independently existing reality nor in the experience of an individual user/agent---makes easier to catch a glimpse of what happens or lies behind the far boundary of that world, which is what I will now attempt. 

A general remark, to begin with. While a junior-level classical mechanics course devotes a considerable amount of time to different formulations of classical mechanics (such as Newtonian, Lagrangian, Hamiltonian, least action), even graduate-level quantum mechanics courses emphasize the wave-function formulation almost to the exclusion of all variants (Heisenberg's matrix formulation, Feynman's path-integral formulation, Wigner's phase-space formulation, etc.).\cite{Styeretal} One disastrous consequence of this state of affairs is that a quantum state's dependence on time has come to be regarded as the time-dependence of an evolving physical state, rather than as a dependence on the time of the measurement to the possible outcomes of which a quantum state serves to assign probabilities.%
\footnote{As it is standard practice to consider objective only those features of a physical situation that do not depend on a particular inertial reference frame, so it ought to be standard practice to extract the physical import of the mathematical apparatus of quantum mechanics from what is common to all formulations of the theory. What is common to all formulations is that they afford tools for calculating statistical correlations between measurement outcomes.}
Another disastrous consequence is the eigenvalue--eigenstate link, an {interpretive principle} that is regarded by many as an essential ingredient of the standard formulation of quantum mechanics.\cite{Gilton2016}  According to it, if a system is ``in'' an eigenstate of the operator $\mathbf{\hat V}$ associated with an observable~$V$, then the value of $V$ is the eigenvalue corresponding to that eigenstate.

The formulation of quantum mechanics best suited to resist the allure of the eigenvalue--eigenstate link is Feynman's, which is based on his formulation of the uncertainty principle\cite{Feynman1948,FHS}:
\bq
\emph{Any determination of the alternative taken by a process capable of following more than one alternative destroys the interference between alternatives.}
\eq
Let us unpack this statement, taking into account that the mere possibility of determining the alternative taken is enough to ``destroy'' the interference between alternatives:
\begin{description}
\item[{Premise A}.]Quantum mechanics provides us with algorithms for assigning probabilities to possible measurement outcomes on the basis of actual outcomes. Probabilities are calculated by summing over alternatives. Alternatives are possible sequences of measurement outcomes. Associated with each alternative is a complex number called ``amplitude.''
\item[{Premise B}.]To calculate the probability of a particular outcome of a measurement~$M_2$, given the actual outcome of a  measurement~$M_1$, choose any sequence of intermediate measurements, and apply the appropriate Rule.
\item[{Rule A}.]If the intermediate measurements are made (or if it is possible to infer from other measurements what their outcomes would have been if they had been made), first square the absolute values of the amplitudes associated with the alternatives and then add the results.
\item[{Rule B}.]If the intermediate measurements are not made (and if it is not possible to infer from other measurements what their outcomes would have been), first add the amplitudes associated with the alternatives and then square the absolute value of the result.
\end{description}
From the wave-function point of view, Rule~B seems uncontroversial. Superpositions are ``normal,'' and what is normal does not call for explanation. What calls for explanation is the existence of mixtures that admit of an ignorance interpretation. From Feynman's point of view, the uncontroversial rule is Rule~A, inasmuch as classical probability theory leads us to expect it. What calls for explanation is why we have to add amplitudes, rather than probabilities, whenever the conditions stipulated by Rule~B are met. 

So why do we have to add amplitudes whenever the conditions stipulated by Rule~B are met? Bohr, with his insistence on the contextuality of the properties that are attributable to quantum systems, came tantalizingly close to providing the answer: 
\bq
\emph{Whenever quantum mechanics instructs us to use Rule~B, the distinctions we make between the alternatives cannot be objectivized. They correspond to nothing in our common external world.}
\eq
If a detector is needed to objectivize a region $D$ of space, to make the property of being in $D$ available for attribution to a quantum system, objective space---the arena of our common external world---cannot be conceived as {intrinsically} differentiated or partitioned. If an electron launched in front of a plate with two slits is detected behind the plate, it can be said to have passed through the union $L\cup R$ of the slits, but if nothing indicates the slit taken by the electron, it cannot be said to have passed through either $L$ or~$R$. The distinction between ``the electron went through~$L$'' and ``the electron went through~$R$'' cannot be objectivized nor, therefore, can the distinction between $L$ and~$R$.

What is more, not only the properties of quantum systems but the systems themselves are defined by the experimental conditions in which they are observed. In other words, quantum systems are \emph{individuated} by these conditions. Particles in particular---``bundles of properties which repeatedly appear together''---``are only individuated by the experimental apparatus in which they are measured or the concrete quantum phenomenon to which they belong''.\cite{Falk2007b} At the same time the \textit{identity of indiscernibles}, a principle according to which two things $A$~and $B$ are one and same thing just in case there is no difference between $A$ and~$B$, requires us to think of particles of the same type as identical in the strong sense of \emph{numerical identity}.%
\footnote{In his Nobel Lecture  on December 11, 1965, Feynman recalled: ``I received a telephone call one day at the graduate college at Princeton from Professor Wheeler, in which he said, `Feynman, I know why all electrons have the same charge and the same mass.' `Why?' `Because, they are all the same electron!'\,''}
What is more, there is no compelling reason to believe that this {numerical} identity ceases when it ceases to have observable consequences owing to the presence of properties by which particles can be distinguished and re-identified. We are free to take the view that what presents itself here and now with these properties and what presents itself there and then with those properties is one and the same entity. I shall refer to this entity as UR, for ``ultimate (or unspeakable) reality'' and in allusion to the fact that the German prefix ``ur-'' carries the sense of ``original'' (as in \emph{Ursprung}, \emph{Ursache}, \emph{etc}.).

\section{UR}\label{sec.UR}
{\leftskip\parindent\small \emph{Omnibus ex nihilo ducendis sufficit unum.}---Gottfried Wilhelm Leibniz\cite{Leibniz}\par
\smallskip\noindent\emph{Felix qui potuit rerum cognoscere causas.}---Virgil\cite{Virgil} \par}\medskip

\noindent In ancient and medieval philosophy, to \emph{be} was either to be a {substance} or to be a property of a substance. Substance was self-existent; everything else depended on a substance for its existence. With Ren\'e Descartes, the human conscious subject assumed the role of substance: to \emph{be} became either to be a subject or to exist as a representation for a subject. Our consciousness, {Descartes} found himself forced to conclude,  is among the world's great certainties. I may not be certain that I am awake now and not dreaming, or I may entertain rational doubts about whether I am really sitting at a desk writing, but I cannot feel this kind of insecurity about whether I am currently having experiences. The world may not exist for all I know, but my {experience}s certainly do. We may harbor any number of illusions, but it takes consciousness for there to be illusions. Nothing attests to the historical importance of Descartes so much as the regularity with which he is still attacked and refuted, more than three hundred years after he wrote. A philosopher who must be refuted so many times must have gotten something profoundly right. Genuinely refutable doctrines need to be refuted only once. Though Descartes nowadays mostly appears in some cartoon version, like {Gilbert Ryle}'s\cite{Ryle49} ``{ghost in the machine}'' or  {Daniel Dennett}'s\cite{Dennett91} ``{Cartesian theater},'' he is absolutely indispensable to the articulation of theories of consciousness like theirs.\cite{Carr99}

But are there not objects over and above those that exist as representations for a subject?  The idea that such objects exist is known as the {representative theory of perception}. It at once raises the question of how the relation between real objects and their representations is to be conceived. Before Kant, there appears to have been no philosopher who did not think of the relation of mental representations to the real world as a relation of similarity. Yet while it seemed to pose no difficulty to think of perceived sizes and shapes as similar to real sizes and shapes, the notion that a perceived color should be similar to an unperceived color in the real world was a more questionable proposition. John Locke and Descartes thus came to distinguish between ``primary qualities,'' which were independent of the perceiving subject, and ``secondary qualities,'' which bore no similarities to sensations but had the power to produce sensations in the perceiving subject. In the end, however, thinking of perceived sizes and shapes as similar to real sizes and shapes proved to be no less questionable than the proposition that color sensations are similar to colors in the external world. George Berkeley made it clear that to ask whether a table is the same size and shape as my mental image of it was to ask an absurd question.

For Kant, \emph{all} qualities were secondary. Everything we say about an object is of the form: it has the power to affect us in such a way. Nothing of what we say about an object describes the object as it is in itself, independently of how it affects us. But Kant did not stop at saying that if I see a desk, there is a thing-in-itself that has the power to appear as a desk, and if I see a chair, there is \emph{another} thing-in-itself that has the power to appear as a chair. For Kant, there was only \emph{one} {thing-in-itself}, the {world-in-itself}, an unspeakable reality that affects us in such a way that we have the sensations that we do and are able to construct, out of the orderly relations among our sensations and with the help of the logical or grammatical relations of our thought or language, our common external world.

Nothing (to my mind) stands in the way of identifying the unspeakable reality of the QBist, which responds to our experimental probings, with the unspeakable reality of Kant, which affects us as described. And there are more ways in which this reality may be related to human experience. As was seen in Sec.~\ref{sec.beyondQBB}, if we objectivize the spatial continuum of our experience---and by ``continuum'' I do not mean some transfinite manifold---then we must think of it as undifferentiated, for the spatial differentiation of our common external world supervenes on its material content, and this does not admit of being objectively partitioned ``all the way down.'' In a sense, therefore, it may be said that ultimately there is only one place, and this is everywhere. And if we take the view that what presents itself here and now with these properties and what presents itself there and then with those properties is one and the same entity, then it may be said that ultimately there is only one ``thing,'' and this {is} everything. It does not seem far-fetched to consider this all-constituting ``thing'' and that all-containing place as further aspects of the unspeakable reality of Kant and the QBists.

What light does quantum mechanics throw on the relation between this unspeakable reality, UR, and our common external world? In the classical limit, quantum states (represented by rays in a Hilbert space) degenerate into classical states (represented by points in a phase space), and the laws of quantum mechanics (which correlate contextual properties {probabilistically}) degenerate into the laws of classical mechanics (which correlate non-contextual properties {deterministically}). In other words, they degenerate into the laws of a world of interacting, causally evolving, and re-identifiable objects. What could this mean in ontological terms? In what way might the quantum laws contribute to or be responsible for the existence, or the coming into being, of such a world---or for the {appearance} of such a world?

In his \emph{Metaphysical Foundations of Natural Science},\cite{KantMFNS} Kant sought to show that certain general physical laws are preconditions for the possibility of experience (in the sense of empirical knowledge). In his \emph{Critique}, he argued that empirical knowledge (i.e., knowledge of an objective world) is possible only if the manifold of appearances allows itself to be sorted into interacting, causally evolving, and re-identifiable objects. More recently, Carl Friedrich von Weizs\"acker,\cite{vW1} a student of Bohr and Heisenberg, has put forth the conjecture---``as a challenge to prove or disprove it''---that \emph{all} physical laws are nothing but preconditions for the possibility of experience, and thus for the existence of a world of interacting, causally evolving, and re-identifiable objects.

If we make room for nontrivial probabilities as indicated in Sec.~\ref{sec.cq}, and if we assume that the objects making up such a world (i)~have spatial extent (they ``occupy space''), (ii)~are sufficiently stable (they neither implode nor explode as soon as they are formed), and (iii)~are ``made up'' of finite numbers of ``things'' that lack spatial extent, then it is possible to derive the quantum laws from the existence of such a world, and thus to show that they are necessary for the existence of our world.\cite{Mohrhoff-QMexplained,Mohrhoff-justso} And if we identify these ``things'' (which are commonly known as ``fundamental particles'' and generally regarded as pointlike) with the one thing that is everything (UR), we can see how the quantum laws contribute to or are responsible for the existence, or the coming into being, of such a world. They describe, in the only way it can be described, the \emph{manifestation} of our common external world.

\section{Particles, atoms, molecules}\label{sec.pa}
{\leftskip\parindent\small It seems clear that quantum mechanics is fundamentally about atoms and electrons, quarks and strings, not those particular macroscopic regularities associated with what we call \emph{measurements} of the properties of these things. But if these entities are not somehow identified with the wave function itself---and if talk of them is not merely shorthand for elaborate statements about measurements---then where are they to be found in the quantum description?---Sheldon Goldstein\cite{Goldstein2017}\par}\medskip

\noindent Now, who or what is responsible for the structure of our shared external world? We, because we have constructed it, or UR, because it has manifested or is manifesting it? The answer I propose is that these alternatives are not exclusive. UR does not simply manifest the world; it manifests the world to itself. And this means that it manifests the world to us, for each one of us is UR, just as each particle is UR. The sole ultimately existing object is also the sole ultimately existing subject, and both UR qua ultimate subject and UR qua ultimate object contribute to the manifestation of our common external world. If we assume that UR simply manifests the world, or that the world exists out of relation to consciousness, we are at a loss to explain how subjects can exist. If we assume that the world is merely the objectivizable intersection of our separate spheres of consciousness, we are at a loss to explain why brains are or appear to be necessary for conscious experience. And if we assume that anything in an experienced world can give rise to the experience of that world, we engender the ``pandemonium of disastrous logical consequences'' alluded to by Schr\"odinger in Sec.~\ref{sec_nonlocal}.%
\footnote{For more on this subject, which obviously exceeds the scope of the present paper, see Chaps.~26 and 27 of~\cite{TWATQM}.}

So how does UR manifest our common external world and, specifically, the {shapes} of the things that exist in it? Shapes resolve themselves into more or less definite, reflexive spatial relations. ``Reflexive'' means that they are self-relations, relations between UR and itself. Spatial relations (relative positions and relative orientations) owe their spatial character to the all-containing yet intrinsically undifferentiated aspect of UR---that sole ultimately existing point which is everywhere. Thus: by entering into (or entertaining) reflexive spatial relations, UR creates (or sustains) the shapes of all the beasts and baubles in the world.

The ultimate (numerically identical) bearers of these relations are often said to be pointlike, and this is taken by many to be a literal truth. What is actually meant, of course, is that the so-called ``ultimate constituents of matter'' lack internal structure. While this, by itself, \emph{may} be consistent with a pointlike form, there are compelling reasons that it implies the absence of form. A fundamental particle then lacks a form because because it lacks internal spatial relations, just as the universe lacks a position because it lacks external spatial relations.

The form defined by a set of spatial relations is an abstract concept. The form of a bipartite object consists of a single relative position~$\mathbf r$. It can be described by a probability distribution over the possible outcomes of a measurement of~$\mathbf r$. The form of an object with $N>2$ components consists of $N(N{-}1)/2$ spatial relations. While the positions of $N{-}1$ components relative to the remaining component can be described with the help of a single probability distribution over a $3(N{-}1)$-dimensional abstract space, the relative positions between these $N{-}1$ components can only be described in terms of correlations between the outcomes of measurements of their positions. The abstractly defined forms of nucleons, nuclei, atoms, and molecules ``exist'' in probability spaces of increasingly higher dimensions. An entirely different kind of form comes into being at the molecular level of complexity: 3-dimensional forms that can be visualized, not as distributions over some probability space, but \emph{as they are}, to wit, the atomic configurations of molecules.

The manifestation of the world (to us) may be described as a transition from the essential unity of UR to the actual {multiplicity} of our common external world. It is brought about by a progressive self-differentiation of UR and a concomitant progressive realization of definiteness (of properties) and distinguishability (of things and regions of space). There is a stage at which UR presents itself as a multitude of formless particles. This stage is probed by high-energy physics and known to us through {correlations} between {detector clicks} (i.e., in terms of transition probabilities between in-states and out-states). There are stages that mark the emergence of form, albeit a type of form that cannot yet be visualized. At energies low enough for atoms to be stable, it becomes possible to conceive of objects with fixed numbers of components, and these we describe in terms of {correlations} between the possible outcomes of measurements. The next stage---closest to the manifested world---contains the first objects with forms that can be visualized, but it is only the final stage---the world manifested to us---that contains the actual {detector clicks} and the actual measurement outcomes on which the properties of atoms and subatomic particles supervene.

How are the intermediate stages to be described---the stages at which the differentiation is incomplete and distinguishability is only partially realized? The answer is that whatever is not completely distinguishable or definite, can only be described by assigning probabilities to what is completely distinguishable or definite, i.e., to the possible outcomes of a measurement. What is instrumental in the manifestation of the world can only be described in terms of {correlations} between events that happen (or could happen) in the manifested world. This, after all, is why the general theoretical framework of contemporary physics is a probability calculus, and why the probabilities are assigned to measurement outcomes. Quantum mechanics affords us a glimpse ``behind'' the manifested world at formless and numerically identical particles, non-visualizable atoms, and partly visualizable molecules, which mark the stages of the transition from the unity of UR to the multiplicity of the experienced world. Instead of being constituent parts of our common external world, atoms, electrons, and quarks (never mind strings) are structures that are instrumental in the {manifestation} of our common external world.

\section{Quantum mechanics and the now}\label{sec_now}
Classical physics ``knows'' about spatial and temporal relations. It also knows about Lorentz-invariant spatiotemporal relations, and it knows how to separate them into frame-dependent spatial and temporal relations. But it knows nothing of absolute positions and absolute times. It has no absolute (as against relative) concepts of ``here'' and ``now.'' Mermin\cite{Mermin_Nature} has expressed the hope that the QBist conversation can be broadened to provide a solution to the problem of the now. How, he asks, can there be no place in physics for something as obvious as the experiential now? So did Einstein. Reporting on a conversation with Einstein, the philosopher Rudolf Carnap\cite{Carnap} wrote:
\bq
 Einstein \dots\ explained that the experience of the Now means something special for man, something essentially different from the past and the future, but that this important difference does not and cannot occur within physics. That this experience cannot be grasped by science seemed to him a matter of painful but inevitable resignation\dots. We both agreed that this was not a question of a defect for which science could be blamed, as Bergson thought\dots. But I definitely had the impression that Einstein's thinking on this point involved a lack of distinction between experience and knowledge.
\eq
Physics concerns the {objectivizable} aspects of our experience. Temporal relations are objectivizable. The experiential now is not. So what? Probabilities, too, cannot be objectivized, as QBists would be the first to insist. The interpretation outlined in the previous sections, on the other hand, makes it possible to objectivize that now which encompasses all our experiences. As there are limits to the objectivation of spatial distinctions, so there are limits to the objectivation of temporal distinctions, and this not only because of the relativistic interdependence of distances and durations.\cite{Mohrhoff_Manifesting} Just as objective space cannot be an intrinsically and completely differentiated spatial expanse, so this now, objectivized, cannot be an intrinsically and completely differentiated temporal expanse.

What Mermin\cite{Mermin_Nature} wants to objectivize, however, is not this experience-encompassing now but a point where consciousness is positioned ``absolutely along the world-line of the being that possesses it.'' The notion that the now of a conscious being is something that moves along a world-line became widely known through a statement by Hermann Weyl\cite{Weyl1949}: ``The objective world simply \emph{is}; it does not \emph{happen}. Only to the gaze of my consciousness, crawling upward along the life line of my body, does a section of this world come to life as a fleeting image in space which continuously changes in time.'' I have remarked on this incoherent notion in several papers. The following comment is from \cite{18errors}.
\bq
If we conceive of temporal relations, we conceive of the corresponding relata simultaneously---they exist at the same time \emph{in our minds}---even though they happen or obtain at different times in the objective world. Since we can't help it, that has to be OK. But it is definitely not OK if we sneak into our simultaneous spatial mental picture of a spatiotemporal whole anything that advances across this spatiotemporal whole. We cannot mentally represent a spatiotemporal whole as a simultaneous spatial whole and then imagine this simultaneous spatial whole as persisting in time and the present as advancing through it. There is only one time, the fourth dimension of the spatiotemporal whole. There is not another time in which this spatiotemporal whole persists as a spatial whole and in which the present advances, or in which an objective instantaneous state \emph{evolves}. If the present is anywhere in the spatiotemporal whole, it is trivially and vacuously everywhere---or, rather, everywhen.
\eq
No doubt, the apparently dimensionless {now} is a deeply mysterious aspect of our experience. Because in our experience \emph{there} appears to be as real as \emph{here}, we are not (in fact, \emph{not sufficiently}) troubled by the perceiving subject's location in an apparently dimensionless \emph{here}. We eliminate this feature of our experience without qualms by making our description of the world translation-invariant in space. But when we make it translation-invariant in time, we loose the defining feature of (philosophical) presentism---the (apparent) unreality of a past or future \emph{then}. This, however, should not surprise us, for like the color of a Burmese ruby the experience of \emph{change} is a quality that can only be defined by {ostentation}---by drawing attention to something of which we are aware. And the same goes for the experience of spatial extension. Here is how Weyl\cite{Weyl1922} explained ``with what little right mathematics may claim to expose the intuitional nature of space'': 
\bq
Geometry contains no trace of that which makes the {space} of intuition what it is in virtue of its own entirely distinctive qualities which are not shared by ``states of addition-machines'' and ``gas-mixtures'' and ``systems of solutions of linear equations.'' It is left to metaphysics to make this ``comprehensible'' or indeed to show why and in what sense it is incomprehensible. We as mathematicians \dots\ must recognise with humility that our conceptual theories enable us to grasp only one aspect of the nature of space, that which, moreover, is most formal and superficial.
\eq

\section{Summary and assessment}\label{sec.sa}
The following affirmations are central to QBism\cite{CFS2002,FMS2014,Fuchs_Perimeter}:
\ben
\item A \emph{probability} is intrinsically and necessarily a degree of confidence or belief, a subjective estimate. Probability assignments are therefore egocentric, single-user.
\item Probability \emph{theory} is a calculus of consistency, a set of criteria for testing coherence between probability judgments.
\item There are \emph{no external criteria} for declaring a probability judgment right or wrong.
\item While probability judgments are necessarily subjective, they are \emph{not intrinsically measures of ignorance}. The subjectivity of probability assignments is consistent with an \emph{objective indeterminism}.
\item A \emph{measurement} is an action taken by a user (of quantum mechanics) or agent (in a quantum world) to elicit one of a set of possible experiences, the apparatus being an integral part of the user or agent. The actual outcome of the measurement is the personal experience elicited thereby, returned by the quantum world in response to her probing.
\item The formal apparatus of quantum mechanics does not account for the actual events/experiences on the basis of which probabilities are assigned, nor for the possible events/experiences to which probabilities are assigned.
\item Measurements are \emph{generative}; they create the properties or values they indicate.\break
\emph{Comment:} According to QBism the creative act is confined to the experience of the experimenter and of those to whom she communicates her finding (and who believe her). It seems to me that QBism could only gain credibility if it situated the creative act in our common external world. The probability that two sane, healthy users looking at the same apparatus see different outcomes is nil. The outcome experienced by a single user should therefore be deemed objective (i.e., contained in our common world) even if it is as yet unknown to any other user. 
 \item \emph{Quantum theory} is an extension or generalization of (Bayesian) probability theory. Like probability theory, it is normative. It provides each user or agent with a calculus of consistency, a set of criteria for testing coherence between probability judgments.
\item The \emph{Born Rule} is a method of transforming or relating probabilities, of getting new ones from old.
\item  There is \emph{no such thing as a true or real quantum state}. A quantum state assignment is equivalent to an assignment of probabilities to the possible outcomes of a measurement or set of compatible measurements. It is as subjective and user-dependent as making a probability judgment.
\item Because a quantum state assignment depends on the information possessed by the user making the assignment, there are potentially as many quantum states for a given system as there are users.
\item Subjective certainty---the assignment of probability~1 to some event---does not warrant objectivity or factuality. Certainty about what a measurement will reveal is not certainty about what is the case. The reality criterion of Einstein, Podolsky, and Rosen\cite{EPR} and the eigenvalue--eigenstate link\cite{Gilton2016} must therefore be rejected.%
\footnote{According to Fuchs \emph{et al.},\cite{FMS2014} ``[t]hat probability-1 (or probability-0) judgments are still judgments, like any other probability assignments, may be the hardest principle of QBism for physicists to accept.'' Really? How can an increase from probability 0.999999 to probability 1 make the difference between a probability assignment and a statement of fact? As I wrote in one of my first papers,\cite{WQMITTU} ``[e]ven the step from probability 1 to factuality crosses a gulf that quantum mechanics cannot bridge.''}
\item The existence of an external reality, undisclosed in experience, is not denied; only that this reality can be represented by a quantum state.
\item All physical systems are to be treated in the same way, ``including atoms, beam splitters, Stern-Gerlach magnets, preparation devices, measurement apparatuses, all the way to living beings and other agents''.\cite{FS2015} The only exception is the individual user's ``own direct internal awareness of her own private experience''.\cite{FMS2014}.\hfill\break
\emph{Comment:} This is the first item to which I must take exception. If user \emph{A} knows that user \emph{B} has obtained a measurement outcome, then it makes no sense for \emph{A} to assign a coherent superposition to \emph{B}, his system, and his apparatus. If \emph{A} is unaware of the outcome \emph{B} has obtained, she will have to assign an incoherent mixture. \emph{B}'s being aware of his outcome is enough to make it part of our common external world, whether or not anyone else is aware of it. 
\item Quantum correlations refer only to timelike separated events: the acquisition of experiences by a single agent.\hfill\break
\emph{Comment:} The meaning of saying that experiences are separated in time is clear enough: they succeed one another in subjective time. What it would mean to say that experiences are separated in space is less clear, though this could be said of the simultaneous experience of a cat and a dog in different parts of subjective space. In either case, the experiences in question are those of a single user. Quantum correlations, however, exist between events in our common external world, not only between events separated by time but also between events separated by space (Sec.~\ref{sec_nonlocal}).
\een
The QBists' fear of nonlocality is misplaced. The diachronic correlations (between outcomes of measurements performed on the same system at different times) are as inexplicable as the synchronic ones (between outcomes of measurements performed on different systems in spacelike relation). We know as little of a physical process by which an event here and now contributes to determine the probability of a \emph{later} event \emph{here} as we know of a physical process by which an event here and now contributes to determine the probability of a \emph{distant} event \emph{now}. \hbox{$\Psi$-ontologists} should be worried,%
\footnote{Closely related to this worry is the perceived inconsistency of quantum mechanics with the principle of local action, a particularly apt formulation of which has been given by Bryce DeWitt and Neill Graham\cite{DeWittGraham71}: ``physicists are, at bottom, a naive breed, forever trying to come to terms with the `world out there' by methods which, however imaginative and refined, involve in essence the same element of contact as a well-placed kick.''}
but not QBists, for whom measurement outcomes are responses from a world beyond the far boundary of our common external world. Because this world (if it can be called a ``world'') is not given in the pure forms of human experience, it is beyond the reach of the concepts at our disposal.

Quantum mechanics does not explain ``how nature does it.'' The theory only explains---via {conservation laws}---why certain things \emph{will not} happen. This is what one would expect if the force at work in the world were an infinite (unlimited) force operating under self-imposed constraints. In this case one would have no reason to be surprised (or dismayed) by the impossibility of explaining the quantum-mechanical correlation laws in terms of physical mechanisms, for it would be self-contradictory to invoke a physical mechanism to explain the working of such a force. What one would need to know is why it works under constraints, and why these constraints---the laws of physics---have the particular form that they do. And this can and has been done.\cite{Mohrhoff-QMexplained,Mohrhoff-justso}

The nonlocality implied by EPR-Bohm correlations, finally, is merely a salient symptom of a deeper nonlocality---the (atemporal) transition from the unity of UR to the multiplicity of the manifested world. This transition is the nonlocal event \emph{par excellence}. Depending on one's point of view, it is either coextensive with spacetime (i.e., completely delocalized) or ``outside'' of spacetime (i.e., not localized at all). Occurring in an anterior relation to space and time, it is the common cause of all events in space and time and all correlations between them---if we grant that the meaning of causality can be extended beyond the spatiotemporal arena of human experience.

QBism, to conclude, is an important contribution to the philosophy of physics. So are the epistemological reflections of Bohr. The interpretation of quantum mechanics, which at present is mired in futile attempts to reify a calculational tool, has much to gain from discarding its stereotypes about Bohr, and QBism, in spite of QBists' own misconceptions about Bohr, can be a great help in doing just this. Bohr, on the other hand, rightly understood, can help dispel misconceptions about QBism and remove its (actual or perceived) inconsistencies.

\section*{Appendix: History of this work (and beyond)}
My first encounter with QBism took place in 2014 at the ``Berge Fest'' (Centre for Quantum Technologies, National University of Singapore) celebrating the 60th birthday of Berthold-Georg (Berge) Englert. My first reaction to QBism as presented there by Ruediger Schack (and to learning that David Mermin was a convert) was one of shock and dismay. This is reflected in the original version of the present paper, which was submitted to \emph{International Journal of Quantum Information} on Sep 11, 2014. 

On Jan 10, 2017 (!) I received an email from B.-G. Englert, one of the managing editors of IJQI, which began with an apology ``for the unusually long time it took to reach an editorial decision about your submission.'' While the paper was rejected ``in its present form,'' I was ``invited to submit a revised version that is properly responsive to the reviewers' criticism and their suggestions for improvement.'' Meanwhile I had become what Mermin once called himself: ``a QBist in the making''.%
\footnote{arXiv:1301.6551 [quant-ph]}
By then I was preoccupied with the second edition of \emph{The World According to Quantum Mechanics} (Singapore, 2018), one of the first textbooks (if not the first) to contain a sizeable chapter on QBism. The revised manuscript I eventually submitted (on Sep 19, 2018) received three positive reviews, all of which recommended publication. However, the 11-page summary of referee reports closed with the following Editor's Note:
\bq
The third referee notes that IJQI is not the most suitable journal for an essay of this kind, and this observation is correct. The discussion about QBism has little or no bearing on quantum information science. IJQI-D-14-00174 has been with IJQI for five years and, therefore, we'll be happy to continue processing this submission with the expectation that a final, publishable version is available after the next revision. Should the author, however, wish to transfer his paper to a more suitable journal, we'll be happy to assist in the transfer.
\eq
Subsequently Englert clarified:
\bq
when submitting to the other journal, you may want to inform the respective editor of the history of this work, possibly by sharing the reviewer's reports on the previous version(s). On your request, I can contact the reviewers and ask them if they are agreeable that I reveal their identity to the editor of the new journal. Should other issues arise when you submit to the other journal, we'll help if we can.
\eq
I made one attempt, inquiring about a possible submission to \emph{Mind and Matter}, which elicited the following response from Harald Atmanspacher, the Editor-in-Chief:
\bq
thanks for inquiring wrt your IJQI manuscript, which I read with interest. I also had a look into the reviewers' comments that you attached\dots.  If we had to review it as a regular submission to Mind and Matter, I anticipate that our reviewers would raise a number of issues over an above those raised by those of IJQI. This could well entail a considerable period of time in addition to the five years it is already pending, until we will be in a position to make a final decision about publication in Mind and Matter. 

\medskip In view of this I would think that it is in your interest to go ahead with the almost completed review process, or so it seems to me, with IJQI. I think this would also be fair wrt the time and effort that their reviewers did already spend on the manuscript. Their overall evaluation seems to be positive, and it remains unclear to me why IJQI editors so quickly pick up on the idea that the ms might be better suited for publication elsewhere.
\eq
There are two reasons why I gave up at this point. (i)~In order to be able to submit the manuscript to a different journal I had to formally withdraw my submission to IJQI. Going ahead with IJQI was thereby ruled out. (ii)~Meanwhile my understanding of QBism and its incorporation in, or adaptation to, my own interpretation of quantum mechanics has evolved further, and has been the subject of several more recent papers:
\begin{itemize}
\item A Qbist Ontology, 	arXiv:2005.14584 [quant-ph]
\item Niels Bohr, objectivity, and the irreversibility of measurements, \emph{Quantum Stud.: Math. Found.} (2019). https://doi.org/10.1007/s40509-019-00213-6
\item Bohr, QBism, and Beyond, arXiv:1907.11405  [quant-ph]
\item `B' is for Bohr, arXiv:1905.07118  [quant-ph]
\end{itemize}
The reason for replacing the original version of this work with the latest available one, finally, is that by now the former has received several citations, of which the following are known to me:
\begin{flushleft}\begin{itemize}
\item Marchildon, L.: Why I am not a QBist. \emph{Found. Phys.} 45, 754--61 (2015)
\item De Ronde, C.: QBism, FAPP and the Quantum Omelette, arXiv:1608.00548 [quant-ph]
\item Glick, D.: QBism and the Limits of Scientific Realism,\break
http://philsci-archive.pitt.edu/id/eprint/16382
\item Zwirn, H.: Is QBism a Possible Solution to the Conceptual Problems of Quantum Mechanics? (to appear in the forthcoming Oxford Handbook of the History of Interpretations and Foundations of Quantum Mechanics), arXiv:1912.11636 [quant-ph]
\end{itemize}\end{flushleft}
I want it to be known to those consulting this work that my thinking has meanwhile quite significantly evolved.
\end{document}